\documentclass[a4paper,11pt,usegeometry]{scrartcl}
\pdfoutput=1

\usepackage[bottom=120pt]{geometry}

\usepackage{amsmath}
\numberwithin{equation}{section}
\usepackage{amssymb}
\usepackage{mathtools}
\usepackage{slashed}

\usepackage{graphicx}
\usepackage{subcaption}

\usepackage{color}
\usepackage{xspace}
\usepackage[bookmarksopen=true]{hyperref}
\hypersetup{
  colorlinks=true,
  linkcolor=blue,
  urlcolor=blue,
  citecolor=blue,
  linktocpage=true
}

\usepackage[sorting=none,style=numeric-comp]{biblatex}
\usepackage{url}
\renewcommand{\arraystretch}{1.2} 

\allowdisplaybreaks

\DeclareRobustCommand{\ensuremathrm}[1]{\ensuremath{\mathrm{#1}}\xspace}

\DeclareRobustCommand{\ensuremathcal}[1]{\ensuremath{\mathcal{#1}}\xspace}

\newcommand{\M}{\ensuremathcal{M}\xspace}
\newcommand{\qb}{\ensuremath{\bar{q}}\xspace}
\newcommand{\Tr}{\ensuremath{\mathrm{Tr}}}
\newcommand{\MSb}{\ensuremath{\overline{\text{MS}}}\xspace}
\newcommand{\evalu}[2]{#1\Bigr|_{#2}}

\DeclareRobustCommand{\order}[1]{\ensuremath{\mathcal{O}\!\left({#1}\right)}\xspace}
\DeclareRobustCommand{\zA}[1]{\ensuremath{\left\langle{#1}\right\rangle}\xspace}
\DeclareRobustCommand{\zB}[1]{\ensuremath{\left[{#1}\right]}\xspace}
\DeclareRobustCommand{\dv}[2][]{\ensuremath{\frac{\mathrm{d}{#1}}{\mathrm{d}{#2}}}}
\DeclareRobustCommand{\pdv}[2][]{\ensuremath{\frac{\partial {#1}}{\partial {#2}}}}

\DeclarePairedDelimiter\abs{\lvert}{\rvert}%

\newcommand{\keV}{\ensuremathrm{\,keV}\xspace}
\newcommand{\MeV}{\ensuremathrm{\,MeV}\xspace}
\newcommand{\GeV}{\ensuremathrm{\,GeV}\xspace}
\newcommand{\TeV}{\ensuremathrm{\,TeV}\xspace}

\newcommand{\email}[1]{\href{mailto:#1}{#1}}

\begin{document}


\titlehead{\raggedleft LMU-ASC~34/22}

\title{\boldmath \LARGE{$h\to gg$ and $h\to\gamma\gamma$ with
  Anomalous Couplings\\ at Next-to-Leading Order in QCD} \unboldmath}

\date{}

\author{{Gerhard Buchalla}\thanks{\email{gerhard.buchalla@lmu.de}} , {Marius H\"ofer}\thanks{\email{m.hoefer@physik.uni-muenchen.de}} , {Christoph M\"uller-Salditt}\thanks{\email{christoph.mueller1@physik.uni-muenchen.de}}}

\publishers{\normalsize{Ludwig-Maximilians-Universit\"at M\"unchen,
 Fakult\"at f\"ur Physik,\\
 Arnold Sommerfeld Center for Theoretical Physics, D--80333 M\"unchen, Germany}}

\maketitle
\thispagestyle{empty}
\setcounter{page}{0}

\begin{center}
\textbf{\textsf{Abstract}}
\end{center}

\begin{abstract}
  We generalize the next-to-leading order QCD calculations for the
  decay rates of $h\to gg$ and $h\to\gamma\gamma$ to the case of
  anomalous couplings of the Higgs boson.
  We demonstrate how this computation can be done in a consistent
  way within the framework of an electroweak chiral Lagrangian,
  based on a systematic power counting.
  It turns out that no additional coupling parameters arise at
  NLO in QCD beyond those already present at leading order.
  The impact of QCD is large for $h\to gg$ and the uncertainties
  from QCD are significantly reduced at NLO. $h\to\gamma\gamma$ is
  only mildly affected by QCD; here the NLO treatment practically
  eliminates the uncertainties. Consequently, our results will allow
  for an improved determination of anomalous Higgs coup\-lings from
  these processes. The relation of our framework to a treatment
  in Standard Model effective field theory is also discussed.
\end{abstract}

\clearpage

\tableofcontents


\section{Introduction}
\label{sec:intro}

The discovery of the Higgs boson a decade ago has opened the door for
novel tests of the mechanism behind electroweak symmetry breaking.
A promising strategy consists in precisely measuring Higgs-boson
couplings, which might deviate from their Standard Model (SM)
expectation and reveal the presence of new dynamics. Such anomalous
couplings are consistently described in the framework of an effective
field theory (EFT). A well-motivated and useful tool for this purpose
is provided by the electroweak chiral Lagrangian including a light
Higgs boson (EWChL, sometimes also referred to as HEFT),
see \cite{Feruglio:1992wf,Buchalla:2015wfa} and refs. therein.
This non-linear version of the electroweak EFT has the practical
advantage of encoding the anomalous Higgs couplings as the dominant
new-physics effects \cite{Buchalla:2015wfa}. This allows us to focus
on the Higgs-boson properties as our main target, and avoids
a proliferation of parameters.

For a reliable determination of Higgs couplings, QCD corrections
have to be taken into account in the calculation of Higgs-boson
processes. Already in the SM, QCD effects do, in general, have a large
numerical impact on the observables
\cite{Baur:1989cm,Spira:1995rr,Beenakker:2002nc,DelDuca:2004wt,Borowka:2016ypz,Spira:2016ztx}.
Higher-order QCD effects can be combined, in a systematic way, with the
anomalous Higgs couplings described by the EWChL~\cite{Buchalla:2015qju}.
An analysis of this type has been performed in \cite{Buchalla:2018yce}
for the case of Higgs-pair production in gluon fusion.
In the present paper, we generalize the calculation of the decay rates for
$h\to\gamma\gamma$ and $h\to gg$ at next-to-leading order (NLO)
in QCD to include new-physics effects in the form of anomalous couplings.
We demonstrate how this can be achieved in a consistent manner
within the framework of the EWChL.

This paper is organized as follows.
In Section~\ref{sec:eft_kin} we summarize the main properties of the EWChL
as an EFT for Higgs processes and define kinematic variables for later use.
Section~\ref{sec:hgamgam} is devoted to the discussion of $h\to\gamma\gamma$,
where the NLO-QCD effects are relatively simple and of moderate size.
To set the stage for the EFT treatment of $h\to gg$, we review the results
for this process at leading order (LO) in QCD in Section~\ref{sec:hgglo}.
Section~\ref{sec:hggnlo} describes our main results, the computation of
$h\to gg$ in the presence of anomalous couplings and with QCD corrections
at NLO. Phenomenological implications of the NLO results for $h\to gg$ are presented in Section~\ref{sec:pheno}.
In Section~\ref{sec:smeft} we discuss how $h\to\gamma\gamma$ and $h\to gg$
could be treated in Standard Model effective field theory (SMEFT),
as an alternative to the EWChL framework we primarily employ.
We conclude in Section~\ref{sec:concl}.
Some further details and examples are collected in the appendix.
App.~\ref{app:IRsub} defines the subtraction of IR divergences in the NLO rate
of $h\to gg$. App.~\ref{app:NNLOfix} explains in detail the dependence of
the $h\to gg$ rate at NLO on the anomalous couplings in the region where
the rate becomes small due to cancellations. In App.~\ref{app:correlations} we give the LO and NLO correlation matrices for the parametric uncertainties of the $h\to gg$ decay rate.
Finally, App.~\ref{app:toy} illustrates the matching of the local $h\gamma\gamma$ and $hgg$
couplings in the EFT to a UV theory in a few toy model scenarios, with a
particular view on the role of QCD corrections in this context.


\section{EFT Lagrangian and kinematic variables}
\label{sec:eft_kin}

The EWChL at lowest order is given by~\cite{Buchalla:2013rka,Buchalla:2013eza}
\begin{eqnarray}\label{l2}
{\cal L}_2 &=& -\frac{1}{4} G_{\mu\nu}^a G^{a \mu\nu}
-\frac{1}{2}\langle W_{\mu\nu}W^{\mu\nu}\rangle 
-\frac{1}{4} B_{\mu\nu}B^{\mu\nu}
+\frac{v^2}{4}\ \langle D_\mu U^\dagger D^\mu U\rangle \, F(\eta)
\nonumber\\
&&  
+\frac{v^2}{2} \partial_\mu \eta \partial^\mu \eta - V(\eta) 
+\bar\psi i\!\not\!\! D\psi - \bar\psi m(\eta,U)\psi
,
\end{eqnarray}
where $\eta\equiv h/v$ with $h$ the Higgs singlet and $v=246\,\GeV$ the
electroweak scale.
$G^a_{\mu\nu}$, $W^\alpha_{\mu\nu}$ and $B_{\mu\nu}$ are the gauge
field strengths of $SU(3)_C$, $SU(2)_L$ and $U(1)_Y$, respectively.
Here $\langle\ldots\rangle$ denotes the trace over $SU(2)_L$ indices.

The electroweak Goldstone bosons $\varphi^\alpha$ are collected in $U=\exp(2i\varphi/v)$,
where $\varphi=\varphi^\alpha t^\alpha$ and $t^\alpha$ denote the
generators of $SU(2)_L$, normalized as
$\langle t^\alpha t^\beta\rangle=\delta^{\alpha\beta}/2$. 
The covariant derivative of the Goldstone field reads 
\begin{equation}\label{dcovu}
D_\mu U=\partial_\mu U + i g W_\mu U -i g' B_\mu U t^3  
\end{equation}
with $SU(2)_L$ and $U(1)_Y$ gauge couplings $g$ and $g'$, respectively.
All SM fermions are collectively written as
$\psi=(u_i,d_i,\nu_i,e_i)^T$, where $u_i,d_i,\nu_i$ and $e_i$ are Dirac spinors
and $i$ is the generation index. The Yukawa term is then given by
the last term in (\ref{l2}) with
\begin{equation}\label{myukawa}
m(\eta,U)\equiv U {\cal M} (\eta) P_R 
+{\cal M}^\dagger (\eta) U^\dagger P_L
,
\end{equation}
where ${\cal M}$ is the block-diagonal mass matrix
\begin{equation}\label{mmdef}
{\cal M}=\mathrm{diag}({\cal M}_u,{\cal M}_d,{\cal M}_\nu,{\cal M}_e)
\end{equation}
acting on $\psi$. In general, the entries ${\cal M}_f\equiv{\cal M}_f(\eta)$ with $f=\{u,d,\nu,e\}$ are $h$-dependent matrices in generation space.

The Higgs-dependent functions are expanded as
\begin{equation}\label{fvdef}
F(\eta)=1+\sum^\infty_{n=1} F_{n} \eta^n\, ,\qquad 
V(\eta)=v^4\sum^\infty_{n=2} V_{n} \eta^n\, ,\qquad
{\cal M}_f(\eta)=\sum^\infty_{n=0} {\cal M}_{f,n} \eta^n
\end{equation}
so that the fermion masses are given by $m_f={\cal M}_{f,0}$. In comparison with the
SM, the Lagrangian in (\ref{l2}) introduces anomalous couplings in the
Higgs sector, out of which only a restricted subset is usually relevant for
a given application. For instance, we can introduce $c_f={\cal M}_{f,1}/m_f$,
which parametrizes potential deviations from the $h\,\bar{f}\, f$ vertex
in the SM. Similarly, assuming custodial symmetry, an anomalous coupling $c_V\equiv F_1/2$ for the
$h\,W^+\, W^-$ and $h\,Z\, Z$ vertices can be defined~\cite{Buchalla:2015wfa}.

\begin{figure}
\centering
\begin{subfigure}[b]{\textwidth}
    \centering
        \includegraphics[scale=0.7]{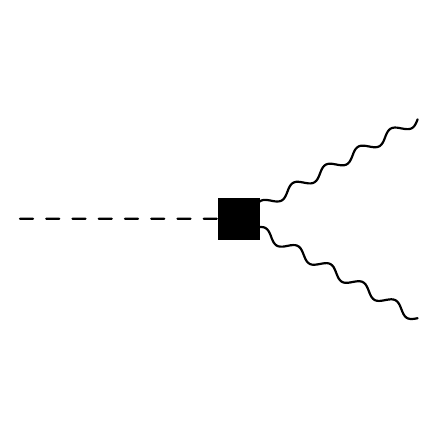}
        \caption{}\label{fig:hgamcpl}
    \end{subfigure}

    \vspace{10pt}

    \begin{subfigure}[b]{\textwidth}
    \centering
        \includegraphics[scale=0.7]{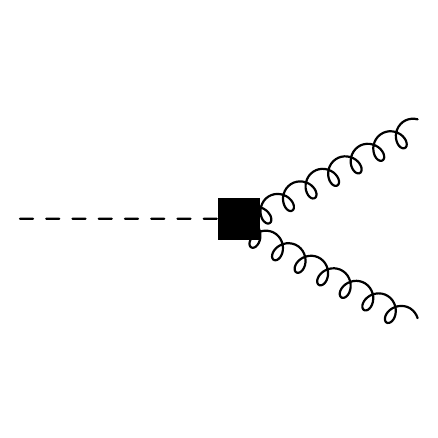}
        ~~~
        \includegraphics[scale=0.7]{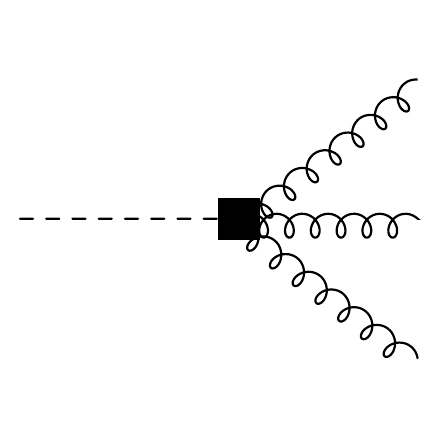}
        ~~~
        \includegraphics[scale=0.7]{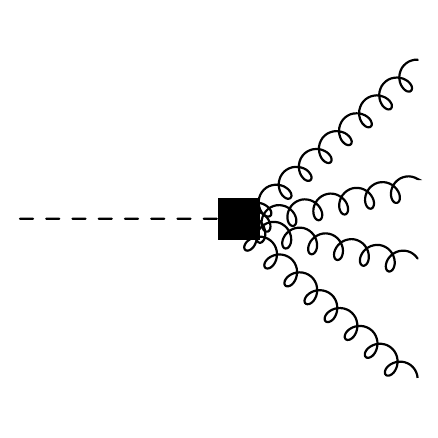}
        \caption{}\label{fig:hgcpl}
    \end{subfigure}
    \caption{New local vertices of the Higgs to photons (a) and gluons (b). The Higgs-photon coupling (a) is generated by the second to last term of (\ref{eq:ewchl}). It is proportional to $c_{\gamma\gamma h}$ and of order $e^2$. The three Higgs-gluon couplings (b) are generated by the last term of (\ref{eq:ewchl}). They are all proportional to $c_{ggh}$ and are of order $g^2_s$, $g^3_s$ and $g^4_s$, respectively}\label{fig:heffcpls}
\end{figure}

Going beyond lowest order in the loop expansion, new terms have to be added
to ${\cal L}_2$. The terms entering at one-loop order in the EWChL are
denoted by ${\cal L}_4$ and can be found in~\cite{Buchalla:2013rka,Sun:2022ssa,Graf:2022rco}.
This introduces further anomalous couplings and also provides the necessary
counterterms for one-loop diagrams from (\ref{l2}).
For our purpose, it is sufficient to focus only on new local interactions
between the Higgs boson and the massless gauge bosons with couplings
$c_{\gamma\gamma h}$ and $c_{ggh}$, respectively, see Figure \ref{fig:heffcpls}.

To summarize, the CP-even terms from the effective Lagrangian
${\cal L}_{eff}\equiv {\cal L}_2 + {\cal L}_4$ with anomalous couplings
relevant for the Higgs decays to two photons or gluons read
\begin{align}
  {\cal L}_{eff}\supset \, & 2c_V\frac{h}{v}\left(m_W^2W_\mu^+W^{-\mu}+
    \frac{1}{2}m_Z^2Z_\mu Z^\mu\right)
  -\sum_f m_f c_f\frac{h}{v} \, \bar{f} \, f \notag\\
 & + \frac{\alpha}{8\pi}c_{\gamma\gamma h}\frac{h}{v}F_{\mu\nu}F^{\mu\nu} +
  \frac{\alpha_s}{8\pi}  c_{ggh} \frac{h}{v}\, G^a_{\mu \nu} G^{a \mu \nu}\, ,
\label{eq:ewchl}
\end{align}
where $\alpha=e^2/4\pi$ and $\alpha_s=g^2_s/4\pi$ are the electromagnetic
and strong fine structure constants, respectively. As stated above, the
anomalous couplings $c_f$ and $c_V$ arise from ${\cal L}_2$
and are leading-order effects in the EFT counting, whereas the local Higgs-gluon and Higgs-photon couplings
$c_{\gamma\gamma h}$ and $c_{ggh}$ are introduced by ${\cal L}_4$ and thus enter at NLO (one-loop) order.
In the SM we have $c_f=c_V=1$ and $c_{\gamma\gamma h}=c_{ggh}=0$.
However, all couplings may have arbitrary values of ${\cal O}(1)$ in general.
In the following, we will neglect the couplings of the Higgs to the first
two lepton generations as well as to up-, down- and strange-quarks due to
their small masses.

We conclude this section by defining some kinematical variables, which
are useful for presenting the rates of $h\to\gamma\gamma$ and $h\to gg$, namely
\begin{align}
    \tau_i &= \frac{m_h^2}{4m_i^2} + i0^+\label{eq:tauidef}
\end{align}
and
\begin{align}
    x_i &= \frac{\sqrt{1-\tau_i^{-1}}-1}{\sqrt{1-\tau_i^{-1}}+1} + i0^+\, ,\label{eq:xidef}
\end{align}
where, for the purpose of analytic continuation, we always assume a small positive imaginary part. In the SM, the leading order decay of Higgs to both photons and gluons is loop induced; $m_i$ in (\ref{eq:tauidef}) usually denotes the mass of the particle running in the loop. We can then distinguish between configurations above and below the particle pair production threshold:
\begin{align}
    \text{below threshold:} && 0<\tau_i<1 &&& x_i = e^{i\theta_i} \quad (0<\theta_i<\pi)\\
    \text{above threshold:} && 1\leq\tau_i<\infty &&& -1\leq x_i<0\,.
\end{align}


\section{Higgs decay to photons}
\label{sec:hgamgam}

\begin{figure}
\centering
    \includegraphics[scale=0.7]{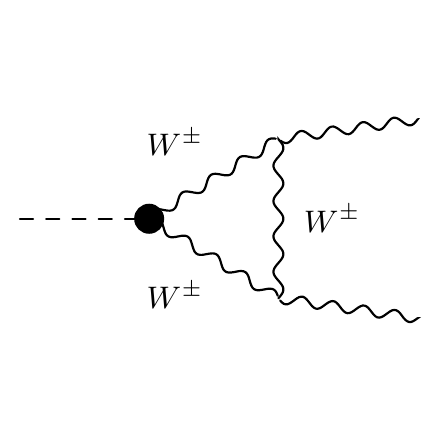}
    ~~~
    \includegraphics[scale=0.7]{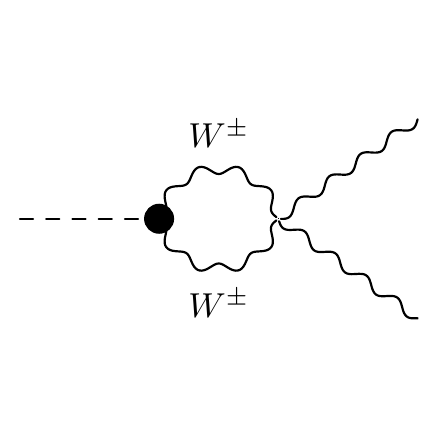}
    ~~~
    \includegraphics[scale=0.7]{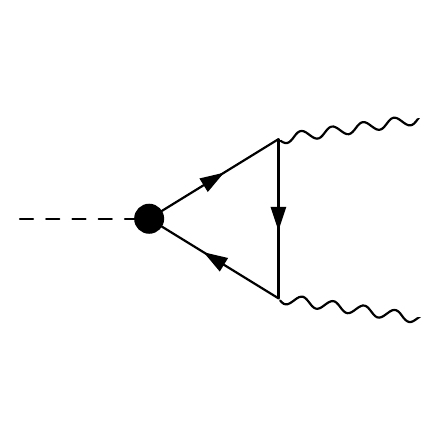}
    ~~~
    \includegraphics[scale=0.7]{figures/h_GG_tree_L4.pdf}
    \caption{Diagrams contributing to the LO, i.e. ${\cal O}(g_s^0)$, amplitudes $A_W$, $A_\ell$, $A_q^{(0)}$ and $A_h$ of the decay $h\to\gamma\gamma$. Here and in the following, black dots and black squares indicate vertices from $\mathcal{L}_2$ and $\mathcal{L}_4$, respectively. The fermion can be both a quark or charged lepton, as long as it is massive. Clockwise and counter-clockwise fermion flow is implicitly understood.}\label{fig:htogamgamLO}
\end{figure}

\begin{figure}
\centering
        \includegraphics[scale=0.7]{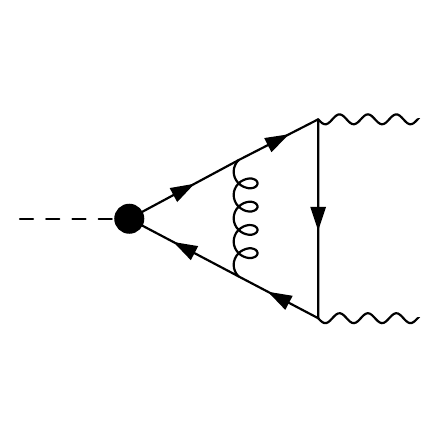}
        ~~~
        \includegraphics[scale=0.7]{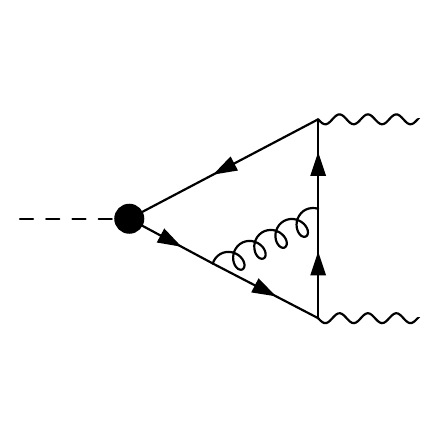}
        ~~~
        \includegraphics[scale=0.7]{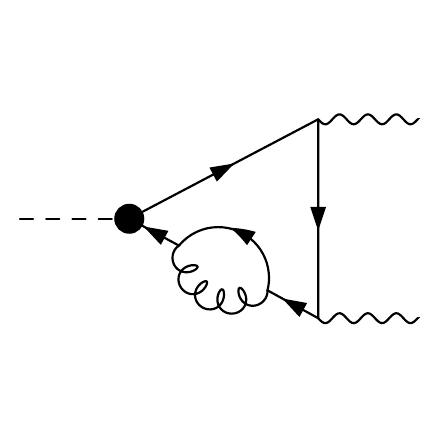}
        ~~~
        \includegraphics[scale=0.7]{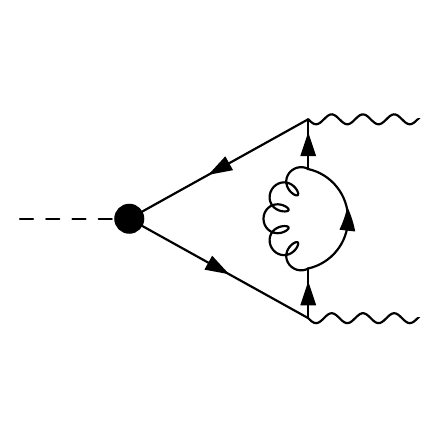}
    \caption{Diagrams contributing to the virtual corrections $A_q^{(1)}$ to the decay $h\to\gamma\gamma$ at ${\cal O}(g_s^2)$. Only the quark-loop diagrams receive QCD corrections at this level.}\label{fig:htogamgamV}
\end{figure}

The decay rate of a Higgs into a pair of photons is given by
\begin{align}
    \Gamma_{h\to\gamma\gamma} &= \frac{\alpha^2}{256\pi^3}\frac{m_h^3}{v^2}\abs*{A_{h\to\gamma\gamma}}^2\,,\label{eq:rategammagamma}
\end{align}
where up to $\order{\alpha_s}$
\begin{align}
    A_{h\to\gamma\gamma} &= c_{\gamma\gamma h}A_h + c_VA_W(\tau_W)+ \sum_\ell c_\ell A_\ell(\tau_\ell) + N_c\sum_qc_qQ_q^2\left(A_q^{(0)}(\tau_q)+\frac{\alpha_s}{4\pi}A_q^{(1)}(\tau_q)\right)\,.\label{eq:amplitudeforgammagamma}
\end{align}
The different subamplitudes $A_i$ describe the coupling of the Higgs to
photons either directly ($A_h$) or through a $W^\pm$ loop ($A_W$),
charged-lepton loop ($A_\ell$) or a quark loop ($A_q^{(0,1)}$), see
Figure~\ref{fig:htogamgamLO}. Only the latter receives QCD corrections at
this perturbative order in $\alpha_s$, see Figure~\ref{fig:htogamgamV}.
The loop functions $A_q$ and their coefficients $c_q$ are
separately renormalization-scale independent.
We have chosen the normalisation of $A_{h\to\gamma\gamma}$ such
that\footnote{Note that the local $h\gamma\gamma$ coupling
  $c_{\gamma\gamma h}$ used here is related to the coupling $c_{\gamma\gamma}$
  defined in \cite{Buchalla:2015qju}
  through $c_{\gamma\gamma h}\equiv 2 c_{\gamma\gamma}$.}
\begin{align}
    A_h &= 1\,.
\end{align}
The coefficient $c_{\gamma\gamma h}$ implicitly contains ${\cal O}(\alpha_s)$
corrections, as will be further discussed in App.~\ref{app:toy}.
The one-loop functions in eq.~(\ref{eq:amplitudeforgammagamma}) are well
known from the SM. They read~\cite{Ellis:1975ap,Shifman:1979eb}
\begin{align}
  A_W(\tau) &= -\frac{2\tau^2+3\tau+3(2\tau-1)f(\tau)}{\tau^2}
   = -\frac{2\left(x^2-8x+1\right)}{(x-1)^2} -\frac{6x(x^2+1)}{(x-1)^4}\ln^2x
              \,,\\[5pt]
  A_\ell(\tau) &= A_q^{(0)}(\tau) = \frac{2(\tau+(\tau-1)f(\tau))}{\tau^2}
    = -\frac{8x}{(x-1)^2} + \frac{2x(x+1)^2}{(x-1)^4}\ln^2x
                 \,,\label{eq:Al0Aq0}
\end{align}
with $x$ defined in (\ref{eq:xidef}) and
\begin{align}
  f(\tau) &= \arcsin^2\sqrt{\tau} = -\frac{1}{4} \ln^2x \,,
\end{align}
which is real valued below the pair production threshold, $0<\tau<1$, and
complex valued otherwise.

The NLO QCD corrections to the quark loop~\cite{Zheng:1990qa,Djouadi:1990aj,Dawson:1992cy,Djouadi:1993ji,Melnikov:1993tj,Inoue:1994jq,Spira:1995rr,Fleischer:2004vb,Harlander:2005rq,Aglietti:2006tp} (Figure~\ref{fig:htogamgamV})
can be decomposed as
\begin{align}
  A_q^{(1)}(\tau) &= A_q^{(1),a}(\tau) -
     6 C_F \tau\pdv[A_q^{(0)}(\tau)]{\tau}X\left(\mu_q^2\right)\,,
\end{align}
where the function $X$ depends on the quark mass renormalization scheme. We have
\begin{align}
    X\left(\mu_q^2\right) &=
        \begin{cases}
            \displaystyle0    &   (\text{OS})\\[5pt]
            \displaystyle\ln\left(\mu_q^2/m_q^2\right) + \frac{4}{3}   &   (\MSb,  \ \text{scheme of \cite{Aglietti:2006tp}})\,,
        \end{cases}\label{eq:Xscheme1}
\end{align}
where $\mu_q$ is the scale at which the mass is renormalized in the case
of a running mass scheme, which is not necessarily identical to the
renormalization scale $\mu_R$ of the strong coupling constant
$\alpha_s$.
The relation between the
quark mass in the on-shell (OS) scheme, $m_\mathrm{OS}$, and the $\MSb$ scheme,
$\bar m(\mu_q)$, to one loop in QCD is
\begin{align}\label{mosmsb}
  m_\mathrm{OS} &= \bar m(\mu_q)\, \left( 1 + \frac{\alpha_s(\bar m)}{\pi}
  \left(\ln\frac{\mu^2_q}{\bar m^2}+\frac{4}{3}\right)\right)\,.
\end{align}  
We remark that in~\cite{Harlander:2005rq} the running
mass is defined as in eq.~(5) of~\cite{Spira:1995rr}, which is different
from the \MSb-mass. This alternative definition corresponds to
\begin{align}
    X\left(\mu_q^2\right) &= \ln\left(\mu_q^2/m_q^2\right) \quad (\text{scheme of~\cite{Spira:1995rr,Harlander:2005rq}})\,.\label{eq:Xscheme2}
\end{align}
The remainder $A_q^{(1),a}$ of the two-loop function can be found in the
literature, see Table~\ref{tab:hgamgamrefs}.
\begin{table}[t]
    \centering
    \begin{tabular}{l|c|c|c|c}
        ref. & $A_W$ & $A_\ell=A_q^{(0)}$ & $A_q^{(1),a}$ & $-6C_F \tau\pdv[A_q^{(0)}]{\tau}$ \\
   \hline
     \cite{Harlander:2005rq,Spira:1995rr} & $A_W^H$ & $\frac{4}{3}F_0^H$ & $4C_FF_0^HC_1^H$ & $4C_FF_0^HC_2^H=2 C_FF_0^HB_2^H$ \\[10pt]
        \cite{Aglietti:2006tp} & $-\mathcal{F}_1^{(1\ell)}$ & $-\mathcal{F}_{1/2}^{(1\ell)}$ & $-4C_F\left(\mathcal{F}_{1/2}^{(2\ell,a)}+\frac{4}{3}\mathcal{F}_{1/2}^{(2\ell,b)}\right)$ & $4C_F\mathcal{F}_{1/2}^{(2\ell,b)}$
    \end{tabular}
    \caption{Loop functions contributing to the decay $h\to\gamma\gamma$ and
      their correspondence in the literature.
      \cite{Harlander:2005rq,Spira:1995rr} set $N_c=3$,
      in~\cite{Aglietti:2006tp} the number of colours is left arbitrary.
      $A_q^{(1),a}$ corresponds to eq.~(10) in~\cite{Fleischer:2004vb},
      which, however, contains typos.
      See footnote~3 of~\cite{Harlander:2005rq}.}
    \label{tab:hgamgamrefs}
\end{table}

Note that the emission of a single gluon off the quark loop ($h\to\gamma\gamma g$) is forbidden by colour symmetry. Therefore, there are no real radiation corrections of $\order{g_s}$ relative to the Born amplitude. As a consequence, the virtual QCD corrections are infrared finite, as any singularities would need to cancel against phase-space singularities in the real corrections by virtue of the KLN theorem~\cite{Kinoshita:1962ur,Lee:1964is}. This enables us to consider an
IR-finite expansion in $g_s$ already at the amplitude
level, given in~(\ref{eq:amplitudeforgammagamma}) to $\order{\alpha_s}$.
The decay rate in~(\ref{eq:rategammagamma})
is then exact to $\order{\alpha_s}$ (NLO QCD) and contains parts of the
$\order{\alpha_s^2}$ (NNLO QCD) corrections. However, to fully capture
NNLO QCD, one would also have to include genuine $\order{\alpha_s^2}$
contributions, that is three-loop diagrams for $h\to\gamma\gamma$, the emission
of two gluons from the quark loop (double real corrections), as well as
two-loop diagrams for $h\to\gamma\gamma$ containing local $hgg$ vertices.
Those contributions are beyond the scope of this work.

We close this section with a brief discussion of the numerical
impact of the NLO QCD effects.
We checked all formulas by independent calculations, except
for the function $A^{(1),a}_q$.
The numbers were obtained with two independent codes. As an additional check we compared with the publicly available program eHDECAY~\cite{Contino:2014aaa}, which implements the results from~\cite{Contino:2013kra}. It contains the $h\to\gamma\gamma$ decay rate as presented in~(\ref{eq:rategammagamma}) and ~(\ref{eq:amplitudeforgammagamma}). Testing several different values of the effective couplings $c_i$ in~(\ref{eq:amplitudeforgammagamma}), we found agreement within the uncertainties of the different implementations for all cases.

Using input parameters from~\cite{PDG:2022}, see also
Table~\ref{tab:input}, we calculate the central values for the various
loop-contributions to the $h\to\gamma\gamma$ amplitude. They are listed in Table~\ref{tab:hgam}.
\begin{table}
\centering
\begin{tabular}{c|r}
    parameter & value \\
    \hline
    $m_h$ & $125.25(17)\GeV$ \\
    $m_t$ (OS mass) & $172.69(30)\GeV$ \\
    $m_b$ (OS mass) & $4.78(6)\GeV$ \\
    $m_c$ (OS mass) & $1.67(7)\GeV$ \\
    $m_\tau$ & $1.77686(12)\GeV$ \\
    $m_W$ & $80.377(12)\GeV$ \\
    $m_Z$ & $91.1876(21)\GeV$ \\
    $\alpha_s(m_Z)$ & $0.1179(9)$ \\
    $G_F$ & $1.1663788(6)\cdot 10^{-5}\GeV^{-2}$ \\
\end{tabular}
\caption{Input parameters for the calculation of the coefficients $A_i$,
  corresponding to the 2022 PDG~\cite{PDG:2022} values. 
  The Higgs vacuum expectation value is derived through its relation to
  the Fermi constant $G_F=(\sqrt{2}v^2)^{-1}$.}\label{tab:input}
\end{table}
\begin{table}[t]
  \centerline{\parbox{14cm}}
\begin{center}
\begin{tabular}{c|c|c|c|c|c}
  $A_W$ & $A_\tau$ & $A_\mu$ & $\frac{4}{3}A_t$ & $\frac{1}{3}A_b$ &
     $\frac{4}{3}A_c$  \\
\hline
$-8.33$ & $-0.024+ 0.022 i$  & $(-3 + i)10^{-4}$  &
    $1.78$ & $-0.027 + 0.023 i$ & $-0.022 + 0.009 i$    \\
\end{tabular}
\end{center}
\caption{Numerical values for the $h\to\gamma\gamma$ amplitude functions
  $A_i$. The quark contributions $A_q$ include the NLO QCD corrections,
  where the quark mass is defined as the pole mass.}
\label{tab:hgam}
\end{table}
These numbers quantify the relative importance of the subamplitudes.
$W$ and top-quark contributions are dominant, $\tau$, $b$ and $c$ loops
only matter when very high precision is required. The lighter fermions
are negligible. Note that here, in contrast to the SM case,
the relative weighting of the subamplitudes is affected by the
anomalous couplings in~(\ref{eq:amplitudeforgammagamma}).

The loops with light fermions have imaginary parts.
Their contribution to the rate is completely negligible:
Assuming SM couplings,
$|A_{h\to\gamma\gamma}|/|\mathrm{Re}\, A_{h\to\gamma\gamma}|$
deviates from unity by less than $10^{-4}$.

We illustrate the impact of QCD corrections on $A_t$,
the dominant contribution from quark loops, using the scheme in~(\ref{eq:Xscheme2}).
For $\mu_t=m_t$ this corresponds to the pole mass.
Using central parameter values and showing the uncertainty
from scale dependence ($m_t/2 < \mu_t < 2 m_t$), we find at LO
and NLO, respectively,
\begin{align}
  A^\mathrm{LO}_t &= 1.3766^{+0.0046}_{-0.0045} \label{atlo} \, ,\\
  A^\mathrm{NLO}_t &= 1.3351^{+0.0000}_{-0.0008} \label{atnlo} \, .
\end{align}
The central value is reduced by about 3 percent at NLO.
At the same time, the small LO uncertainty of 3 permille
is reduced by another order of magnitude at NLO and thus
essentially eliminated.

A convenient analytical expression for the top-quark function
at NLO can be obtained from an expansion in the variable $\tau$.
To linear order in $\tau$ it reads
\begin{align}\label{atexp}
  A_t(\tau) &= \frac{4}{3}+\frac{14}{45} \tau +\frac{\alpha_s}{\pi}
 \left( -\frac{4}{3}+\frac{488}{405} \tau - \frac{28}{45}\tau      
   X\left(\mu_q^2\right)\right)\, ,
\end{align}
which is accurate at the permille level.

For SM couplings we find with our central parameter set 
\begin{align}\label{gamh2pnum}
  \Gamma_{h\to\gamma\gamma} &= 9.54\, \keV \, ,
\end{align}
including the NLO QCD corrections (the LO value is 9.41 keV).
The error from scale dependence
in the $t$, $b$ and $c$ amplitudes is safely below a permille.

Displaying the dependence of the rate on the anomalous couplings
we may write
\begin{align}\label{gamh2pci}
  \Gamma_{h\to\gamma\gamma}/\keV &=
   15.098\,  c_W^2 - 6.451\,  c_t c_W - 3.624\,  c_{\gamma\gamma h} c_W
  +0.774\, c_{\gamma\gamma h} c_t + 0.689\, c_t^2 \nonumber\\
  & + 0.217\, c_{\gamma\gamma h}^2 - 0.012\,  c_b c_{\gamma\gamma h}
    - 0.009\,  c_c c_{\gamma\gamma h}
   - 0.021\,  c_b c_t - 0.017\, c_c c_t\nonumber\\  
  & -0.010\, c_{\gamma\gamma h} c_\tau - 0.018\, c_t c_\tau  
    + 0.097\,  c_b c_W + 0.079\, c_c c_W  + 0.085\,  c_\tau c_W \, .
\end{align}
Here we have dropped terms with coefficients of less than $0.001$.
Again, the NLO QCD uncertainties are negligible.


\section{\boldmath $h\to gg$ at LO in QCD\unboldmath}
\label{sec:hgglo}

\begin{figure}
\centering
    \includegraphics[scale=0.7]{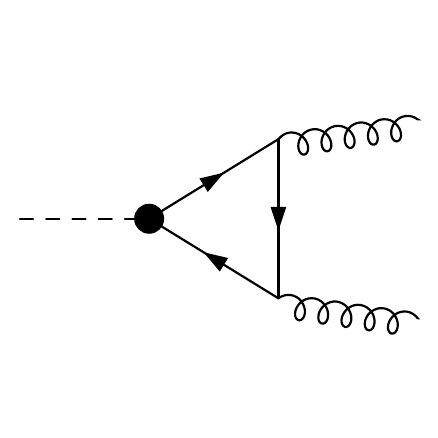}
    ~~~
    \includegraphics[scale=0.7]{figures/h_gg_tree_L4.pdf}
    \caption{Diagrams contributing to the decay $h\to gg$ at LO, both in the chiral counting (one-loop order) and in the QCD coupling (order $g^2_s$).}\label{fig:LO}
\end{figure}

At LO the decay rate of a Higgs into two gluons is given by
\begin{align}
    \Gamma_{h\to gg}^{LO} &= \frac{\alpha_s^2}{256\pi^3}\frac{m_h^3}{v^2}(N_c^2-1)\abs*{A_{h\to gg}^{(0)}}^2\,,
\end{align}
where
\begin{align}
    A_{h\to gg}^{(0)} &= c_{ggh}A_h^{(0)}+\frac{1}{2}\sum_qc_qA_q^{(0)}(\tau_q)\,.\label{eq:Ahgg0}
\end{align}
The first term is the local Higgs-gluon interaction (Figure~\ref{fig:LO} right). As for the $h\to\gamma\gamma$ amplitude, we have chosen the normalisation of $A_{h\to gg}^{(0)}$ such that
\begin{align}
    A_h^{(0)} & = 1\,\label{eq:Ah0}.
\end{align}
The second term of~(\ref{eq:Ahgg0}) accounts for the SM-like contribution from heavy-quark loops (Figure~\ref{fig:LO} left). Note that compared to (\ref{eq:Al0Aq0}), there is a different overall factor of the triangle contribution $A_q^{(0)}(\tau_q)$, stemming from the different colour structures of the quark-gluon and quark-photon vertices\footnote{We have $T_{ij}^a$ for the quark-gluon and $Q_q\delta_{ij}$ for the quark-photon vertex with $T^a$ the $SU(N_c)$ generators in the fundamental representation ($i,j=1,\dots,N_c$) and normalization $\Tr[T^aT^b]=\frac{1}{2}\delta^{ab}$.}. $A_q^{(0)}(\tau_q)$ is the same function which we already encountered in $h\to\gamma\gamma$, see Table~\ref{tab:hgamgamrefs}. Compact tree-level helicity amplitudes for the decay of a scalar particle like the Higgs into an arbitrary number of fermions and gluons via the interaction term $hG_{\mu\nu}^aG^{a\mu\nu}$ can be found in~\cite{Dixon:2004za,Badger:2004ty}.

In order to examine the effect of variations of the effective couplings
$c_{ggh}$ and $c_q$, the decay rate can be expressed as a polynomial bilinear
in the couplings,
\begin{align}
    \Gamma_{h\to gg}^{LO} &= A_{gg}^{LO}c_{ggh}^2 + A_{tt}^{LO}c_t^2 + A_{bb}^{LO}c_b^2 + A_{tg}^{LO}c_{ggh}c_t + A_{bg}^{LO}c_{ggh}c_b + A_{bt}^{LO}c_{t}c_b\,,\label{eq:rateLO}
\end{align}
where we considered only third generation quarks, i.e. $q=b,t$.
The contributions from the other quarks are negligible due to their
small mass and hence their suppressed coupling to the Higgs. Unlike in the $h\to\gamma\gamma$ decay there is no enhancement of the c-quark wrt. the b-quark contribution by electromagnetic charge factors.


\section{\boldmath $h\to gg$ at NLO in QCD\unboldmath}
\label{sec:hggnlo}
At NLO in QCD we have to consider both virtual (V) and real radiation (R) corrections. The former consist of all $\order{g_s^4}$, two-loop order $h\to gg$ diagrams, the latter comprise all $\order{g_s^3}$ one-loop order diagrams with one extra massless coloured particle in the final state, i.e. $h\to ggg$ and $h\to g q\bar q$, where $q$ ($\bar q$) is a massless quark (anti-quark).

The NLO decay rate can be written as
\begin{align}
  \Gamma_{h\to gg}^{NLO} &= \Gamma_{h\to gg}^{LO} + \Gamma_{h\to gg}^V +
    \Gamma_{h\to gg}^R\,.\label{eq:rateNLO_a}
\end{align}
Both the virtual and the real radiation contribution are in fact infrared (IR) divergent. The former due to explicit poles in $\epsilon=(4-D)/2$ from the dimensionally regulated loop integrals ($D$ is the number of space-time dimensions), the latter due to phase-space configurations with soft or collinear partons for which the matrix elements are singular. The singularities cancel in the sum of both contributions and we obtain a finite, physically meaningful result. In practise a suitable IR scheme has to be chosen to deal with the cancellation of the singularities. We adopt the antenna subtraction formalism~\cite{GehrmannDeRidder:2005cm,Daleo:2006xa,Currie:2013vh}.

Similarly to the LO rate in \ref{eq:rateLO}, we write the NLO decay rate as a polynomial
in the effective couplings,
\begin{align}
    \Gamma_{h\to gg}^{NLO} &= A_{gg}^{NLO}c_{ggh}^2 + A_{tt}^{NLO}c_t^2 + A_{bb}^{NLO}c_b^2 + A_{tg}^{NLO}c_{ggh}c_t + A_{bg}^{NLO}c_{ggh}c_b + A_{bt}^{NLO}c_{t}c_b\,.\label{eq:rateNLO}
\end{align}

\subsection{Virtual corrections}
\label{subsec:V}
\begin{figure}
\centering
    \begin{subfigure}[b]{\textwidth}
    \centering
        \includegraphics[scale=0.7]{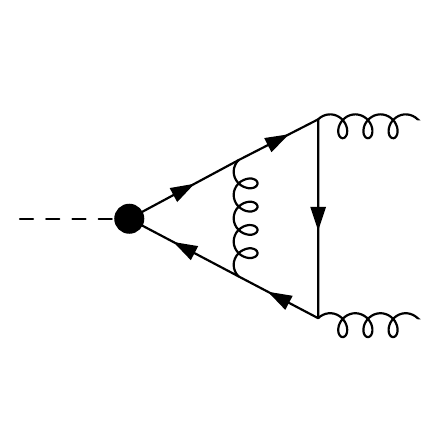}
        ~~~
        \includegraphics[scale=0.7]{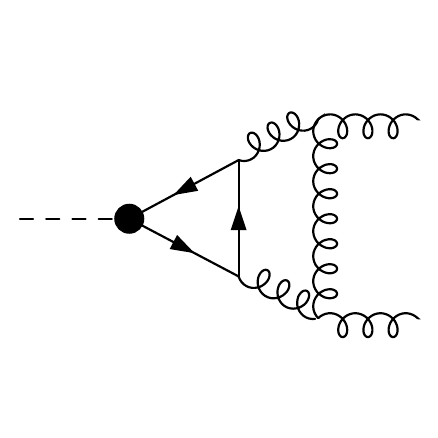}
        ~~~
        \includegraphics[scale=0.7]{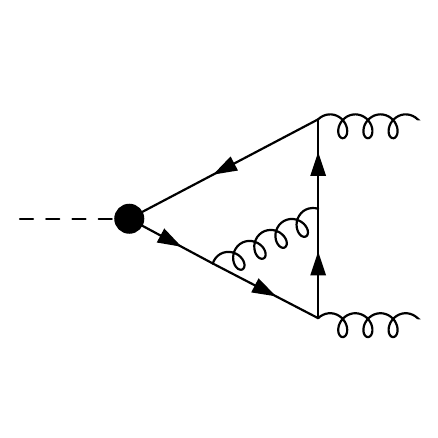}
        ~~~
        \includegraphics[scale=0.7]{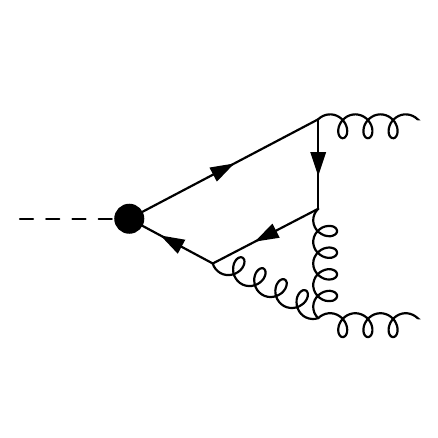}
        ~~~
        \includegraphics[scale=0.7]{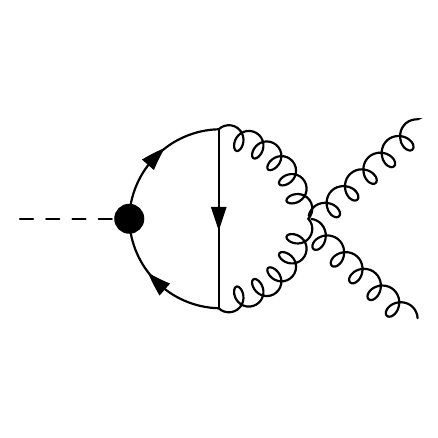}
        ~~~
        \includegraphics[scale=0.7]{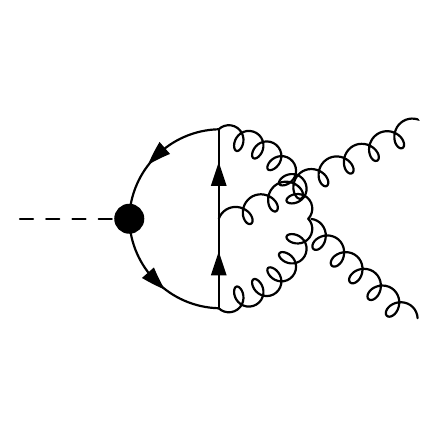}
        ~~~
        \includegraphics[scale=0.7]{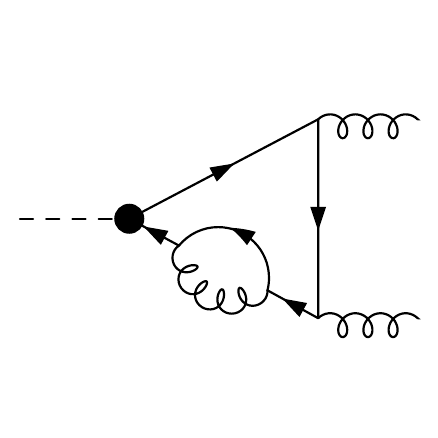}
        ~~~
        \includegraphics[scale=0.7]{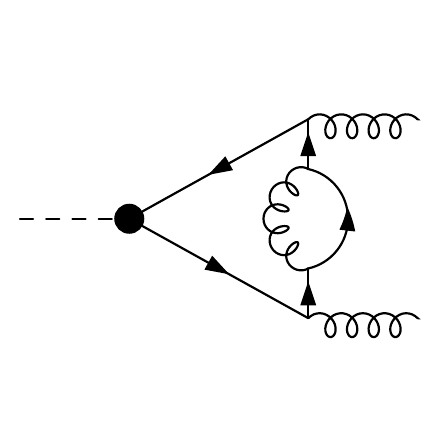}
        \caption{}\label{fig:VL2}
    \end{subfigure}    
   
    \vspace{10pt}
   
    \begin{subfigure}[b]{\textwidth}
    \centering
        \includegraphics[scale=0.7]{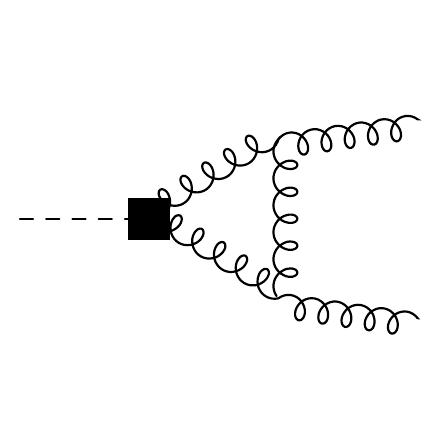}
        ~~~
        \includegraphics[scale=0.7]{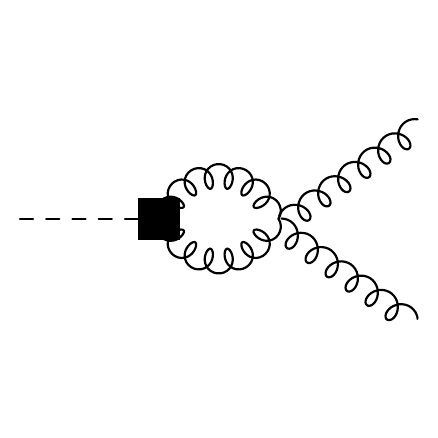}
        \caption{}\label{fig:VL4}
    \end{subfigure} 
    \caption{Diagrams contributing to the virtual corrections to the decay $h\to gg$ at $\order{g_s^4}$. (a) Genuine two-loop diagrams with vertices from $\mathcal{L}_2$ only. (b) one-loop diagrams with one effective vertex from $\mathcal{L}_4$.}\label{fig:V}
\end{figure}

\begin{figure}
\centering
    \begin{subfigure}[b]{0.6\textwidth}
    \centering
        \includegraphics[scale=0.7]{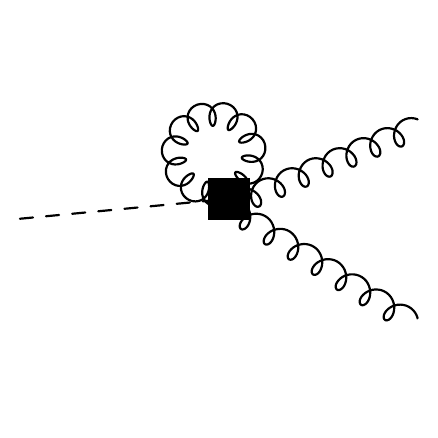}
        ~~~
        \includegraphics[scale=0.7]{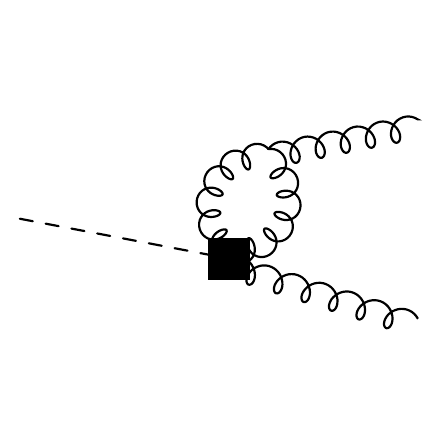}
        \caption{}\label{fig:VL4nc}
    \end{subfigure}  
    \begin{subfigure}[b]{0.3\textwidth}
    \centering
        \includegraphics[scale=0.7]{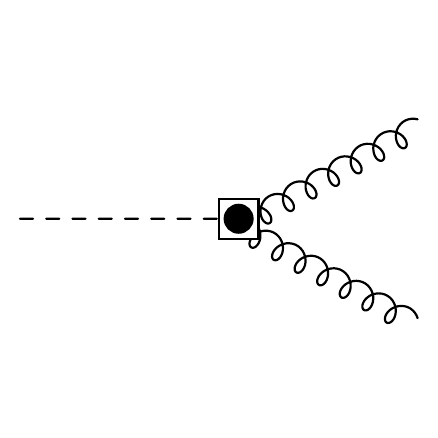}
        \caption{}\label{fig:VL6nc}
    \end{subfigure}    
    \caption{Diagrams of two-loop order not contributing to the virtual corrections of $\order{g_s^4}$. While being of $\order{g_s^4}$, the scaleless diagrams (a) vanish by virtue of dimensional regularization. The diagram (b) contains an effective $hgg$ vertex coming from the NNLO chiral Lagrangian $\mathcal{L}_6$. It gives no additional contribution to the considered order as further discussed in the text.}\label{fig:Vnc}
\end{figure}

There are three distinct classes of diagrams contributing to the virtual
corrections, see Figures~\ref{fig:V} and \ref{fig:Vnc}:
\begin{enumerate}
    \item Genuine two-loop diagrams with vertices from $\mathcal{L}_2$ only.
    \item one-loop diagrams with a single one-loop order effective vertex of
      chiral dimension 4, i.e. from $\mathcal{L}_4$.
    \item One tree level diagram with an effective vertex of
      chiral dimension 6 coming from $\mathcal{L}_6$.
\end{enumerate}
Diagrams of the first class are shown in Figure~\ref{fig:VL2}. Up to the
rescaling by the effective couplings $c_q$ they correspond to the diagrams
needed to calculate the two-loop amplitude in the ordinary SM with
full mass dependence. The second class of diagrams (Figure~\ref{fig:VL4})
associated with $c_{ggh}$ has no correspondence in the SM. While
$\mathcal{L}_4$ also provides $hggg$ and $hgggg$ vertices of the right order
in $g_s$, the relevant one-loop diagrams (Figure~\ref{fig:VL4nc}) vanish when
evaluated in dimensional regularization. Higgs plus multiparton one-loop
amplitudes with local Higgs-gluon interactions have been calculated in~\cite{Berger:2006sh}. At last there is the single tree-level diagram
with an effective vertex of chiral dimension $6$ (Figure~\ref{fig:VL6nc}).
Such a diagram could in principle contribute at the order under
consideration, that is two-loop order in the EWChL and at
${\cal O}(g^4_s)$. However, from gauge invariance any local $hgg$ vertex
can be expressed by an operator $hG^a_{\mu\nu}G^{a\mu\nu}$, which is
identical to the corresponding term already included in ${\cal L}_4$.
In fact, a $d_\chi=6$ operator such as
\begin{align}
  {\cal O}_{6,hgg} = g^2_s D_\rho G_{\mu\nu}^aG^{a\mu\nu}\partial^\rho h \,,
  \label{eq:o6hgg}
\end{align}
or similar terms, can be eliminated using integration by parts and
equations of motion (eom) in favour of the operator $hG^a_{\mu\nu}G^{a\mu\nu}$.
For example,
\begin{align}
  {\cal O}_{6,hgg} &= -\frac{1}{2} g^2_s G_{\mu\nu}^a G^{a\mu\nu} \partial^2 h =
  \frac{1}{2v} g^2_s G_{\mu\nu}^a G^{a\mu\nu} \left( V'-\frac{v^2}{4}
  \langle D_\mu U^\dagger D^\mu U\rangle\, F'+\bar\psi m'\psi \right)
  \nonumber \\
  &= \frac{m^2_h}{2} g^2_s G_{\mu\nu}^a G^{a\mu\nu} h + \ldots \,,
  \label{eq:o6hggr}
\end{align}
where we dropped total derivatives and, in the last step, terms
with additional fields, which do not contribute at the relevant order.
In general, local terms with $d_\chi=6$ for the $hgg$ vertex therefore
correspond to subleading contributions in the coefficient
\begin{align}
c_{ggh} = c^{(0)}_{ggh} +{\cal O}(g^2_s, m^2_h/\Lambda^2)\, ,
\end{align}  
with the leading term $c^{(0)}_{ggh}={\cal O}(1)$.
The $g^2_s$ corrections are part of the NLO QCD effects, as will be
further discussed in App.~\ref{app:toy}.
The terms $\sim m^2_h/\Lambda^2$ are formally negligible at the
considered order. They do not scale as $g^4_s$ and are not part
of the NLO QCD corrections.
In practice, all these effects are implicitly contained in
the coefficient $c_{ggh}$.

Other terms at $d_\chi=6$ contribute only beyond the order
we are considering. For instance, the operator
\begin{align}
  {\cal O}_{6,hgq} = g^2_s D^\mu G_{\mu\nu}^a G^{a\nu\lambda}\partial_\lambda h
  =  g^3_s \bar q\gamma_\nu T^a q\, G^{a\nu\lambda}\partial_\lambda h \,,
  \label{eq:o6hgq}
\end{align}
where we have used the gluon eom in the last step,
can interfere with the diagrams in Figure~\ref{fig:Rgqqb}.
However, this is one loop order higher than the squares
of Figure~\ref{fig:Rgqqb} entering at NLO, and can be consistently
neglected.

The contribution of the virtual corrections to the decay rate is given by
\begin{align}
    \Gamma_{h\to gg}^{V} &= \frac{\alpha_s^3}{512\pi^4}\frac{m_h^3}{v^2}(N_c^2-1)\mathrm{Re}\left\{A_{h\to gg}^{(0)\dagger}A_{h\to gg}^{(1)}\right\}\,,\label{eq:hggV}
\end{align}
where we explicitly pulled out the coupling factor $\alpha_s/4\pi$ from the NLO part of the amplitude as in (\ref{eq:amplitudeforgammagamma}) with
\begin{align}
    A_{h\to gg}^{(1)} &= N_cI(\epsilon)A_{h\to gg}^{(0)}(\epsilon) + \frac{1}{2}\sum_qc_qA_{q,g,\mathrm{fin}}^{(1)}(\tau_q) + \order{\epsilon}\, .
\end{align}  
The renormalization of the amplitude is carried out in the \MSb-scheme, apart from the quark-mass, for which we also considered the OS scheme and the scheme from~\cite{Spira:1995rr,Harlander:2005rq}.
The IR finite part of the amplitude can then be decomposed as
\begin{align}
    A_{q,g,\mathrm{fin}}^{(1)}(\tau) &= A_{q,g}^{(1),a}(\tau)-6C_F\tau\pdv[A_q^{(0)}(\tau)]{\tau}X\left(\mu_q^2\right)\,,\label{eq:Aqgfin1}
\end{align}
where $X(\mu_q^2)$ is defined above, see equations~(\ref{eq:Xscheme1}) and~(\ref{eq:Xscheme2}) and the expression $A_{q,g}^{(1),a}(\tau_q)$ can be found in the literature, see Table~\ref{tab:hggVrefs}. The IR singular behaviour is contained in\footnote{This object is closely related to Catani's one-loop insertion-operator
$\boldsymbol{I}_{ij}^{(1)}(\epsilon)$~\cite{Catani:1998bh} via
\begin{align*}
    I(\epsilon) &= \frac{2}{N_c}\boldsymbol{I}_{gg}^{(1)}(\epsilon)+\frac{\beta_0}{N_c}L+\order{\epsilon}\, .
\end{align*}}
\begin{align}
    I(\epsilon) &= -\frac{2}{\epsilon^2}-\frac{1}{\epsilon}\left(\frac{\beta_0}{N_c}+2L\right)-L^2+\frac{\pi^2}{6}\,,\label{eq:Iepsilon}
\end{align}
where
\begin{align}
    \beta_0 &= \frac{11}{3}N_c-\frac{2}{3}N_F
\end{align}
is the first term of the QCD $\beta$-function and $L=\log\mu_R^2/m_h^2+i\pi$,
with $\mu_R$ the renormalization scale. $I(\epsilon)$ multiplies the LO amplitude, for which we now also have to consider terms up to $\order{\epsilon^2}$,
\begin{align}
    A_{h\to gg}^{(0)}(\epsilon) &= c_{ggh}S_\epsilon^{-1}A_h^{(0)}+\frac{1}{2}\sum_qc_q\left(\frac{m_q^2}{\mu_R^2}\right)^{-\epsilon}A_q^{(0)}(\tau_q,\epsilon)
\end{align}
with $S_\epsilon = (4\pi)^\epsilon e^{-\gamma\epsilon}$ and $\gamma=0.57721\dots$ is the Euler-Mascheroni constant. $A_q^{(0)}(\tau_q,\epsilon)$ corresponds to $M_\mathrm{f}^{(0)}=-\bar{\M}_\mathrm{f}^{(0)}$ from~\cite{Anastasiou:2006hc}. Its zeroth order term is just the LO expression $A_q^{(0)}(\tau_q)$. We work in the $N_F=5$
scheme, i.e. all quarks besides the top are considered as massless,
except for the bottom-quark in the loop contribution to $h\to gg$.

\begin{table}[t]
    \centering
    \begin{tabular}{l|c|c}
            refs. & $A_{q,g}^{(1),a}$ & $-6C_F\tau\pdv[A_q^{(0)}]{\tau}$\\
        \hline
            \cite{Harlander:2005rq,Spira:1995rr} & $\frac{8}{9}C_A\left(F_0^HB_1^H-2F_0^HC_1^H\right)+4C_FF_0^HC_1^H$ & $4C_FF_0^HC_2^H=2C_FF_0^HB_2^H$ \\[10pt]
            \cite{Aglietti:2006tp} & $-4C_A\mathcal{G}_{1/2}^{(2\ell,C_A)}-4C_F\left(\mathcal{F}_{1/2}^{(2\ell,a)}+\frac{4}{3}\mathcal{F}_{1/2}^{(2\ell,b)}\right)$ & $4C_F\mathcal{F}_{1/2}^{(2\ell,b)}$ \\[10pt]
            \cite{Anastasiou:2006hc} & $M_{f,\mathrm{fin}}^{(1)}-6C_Fm_q^2\pdv[M_f^{(0)}]{m_q^2}\left(\frac{4}{3}+\log\frac{m_h^2}{m_q^2}\right)$ & $6C_Fm_q^2\pdv[M_f^{(0)}]{m_q^2}$
    \end{tabular}  
    \caption{References for the two-loop corrections to the quark-loop contribution to the $h\to gg$ rate and the expressions therein, which correspond to the functions introduced in this paper. \cite{Harlander:2005rq,Spira:1995rr} set $N_c=3$, \cite{Aglietti:2006tp,Anastasiou:2006hc} keep the number of colours arbitrary.}
    \label{tab:hggVrefs}
\end{table}

\subsection{Real radiation corrections}
\label{subsec:R}
\begin{figure}
\centering
    \begin{subfigure}[b]{\textwidth}
    \centering
        \includegraphics[scale=0.7]{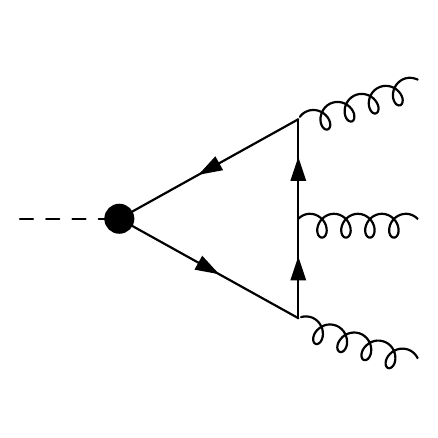}
        ~~~
        \includegraphics[scale=0.7]{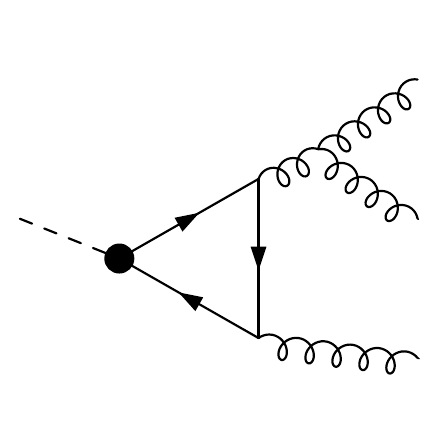}
        ~~~
        \includegraphics[scale=0.7]{figures/h_ggg_tree_L4_1.pdf}
        ~~~
        \includegraphics[scale=0.7]{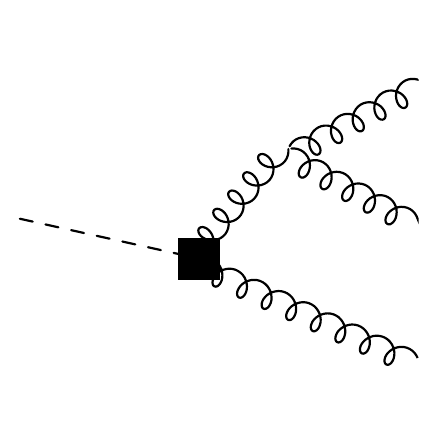}
        \caption{}\label{fig:Rggg}
    \end{subfigure}    
   
    \vspace{10pt}
   
    \begin{subfigure}[b]{\textwidth}
    \centering
        \includegraphics[scale=0.7]{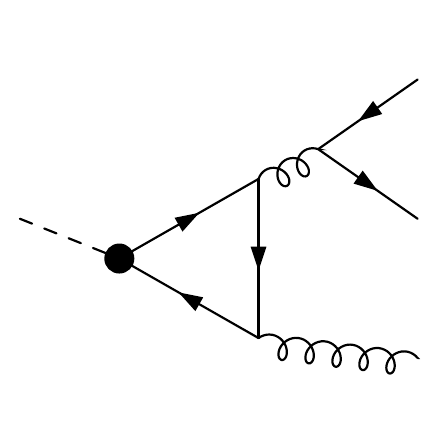}
        ~~~
        \includegraphics[scale=0.7]{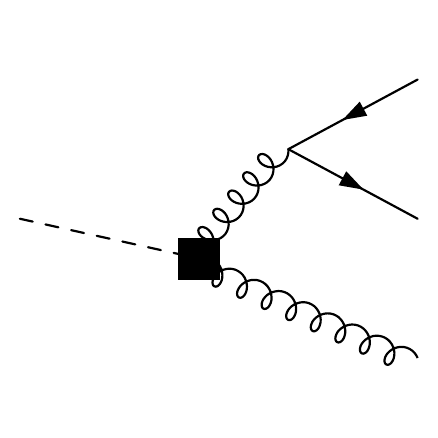}
        \caption{}\label{fig:Rgqqb}
    \end{subfigure} 
    \caption{Example diagrams contributing to the real emission corrections to the decay $h\to gg$ at $\order{g_s^3}$ at the amplitude level. (a) $ggg$ decay channel. (b) $gq\qb$ decay channel.}
    \label{fig:R}
\end{figure}

Moving on to the real corrections, we have to consider diagrams where in comparison to the Born level expression, an additional gluon is radiated into the final state (Figure~\ref{fig:Rggg}). This includes a contribution featuring an effective $hggg$ vertex. Originating from the same term in $\mathcal{L}_4$ as the $hgg$ vertex, it also comes with the coupling $c_{ggh}$. \\
Besides the $h\to ggg$ channel, we have to include the possibility of a gluon splitting into a massless quark-anti-quark pair, i.e. the channel $h\to g(g\to q\qb)$ (Figure~\ref{fig:Rgqqb}). This is because in the collinear configuration, the $q\qb$ pair is indistinguishable from a gluon. As mentioned above, we consider $N_F=5$ massless quark flavours. Note that since the coupling of the Higgs to quarks is directly proportional to the quark mass, there are no diagrams contributing to the $h\to gq\qb$ channel where the Higgs couples directly to the light massless quarks and the final state gluon is radiated off the quark line. The real emission contribution to the decay rate is then simply the sum of both channels,
\begin{align}
    \Gamma_{h\to gg}^R &= \Gamma_{h\to gg}^{R,ggg} + N_F\Gamma_{h\to gg}^{R,gq\qb}\,.
\end{align}

\subsubsection[$h\to ggg$ channel]{\boldmath $h\to ggg$ channel\unboldmath}
\label{subsubsec:Rggg}
The contribution of the $ggg$ channel to the NLO decay rate is given by
\begin{multline}
    \Gamma_{h\to gg}^{R,ggg} = \frac{\alpha_s^3}{24\pi m_hv^2}N_c(N_c^2-1)\\\times\int\mathrm{d}\Phi_3(p_1,p_2,p_3)\sum_\lambda\mathcal{H}^\lambda(s_{12},s_{23},s_{13})\abs*{A_{h\to ggg}^{(0)\lambda}(s_{12},s_{23},s_{13})}^2\,,\label{eq:hggRggg}
\end{multline}
where $s_{ij}=(p_i+p_j)^2$ with $p_i$ and $p_j$ denoting either two of the outgoing gluon momenta (we have $s_{12}+s_{23}+s_{13}=m_h^2$). Adjusted to the case at hand (decay of a scalar particle), the three-body phase space reads
\begin{align}
  \mathrm{d}\Phi_3(p_1,p_2,p_3) &=
     \frac{(2\pi)^{2\epsilon-3}}{2^{4-2\epsilon}\Gamma(2-2\epsilon)}
       (m_h^2)^{\epsilon-1}(s_{12}s_{23}(m_h^2-s_{12}-s_{23}))^{-\epsilon}
                   \  \mathrm{d}s_{12}\,\mathrm{d}s_{23}\,.
\end{align}
The form of the amplitude depends on the concrete helicity configurations $\lambda$ of the three gluons, so the helicity summation has to be left explicit. We factored out the object $\mathcal{H}^\lambda$ which captures the helicity dependence of the tree-level contribution, so that
\begin{align}
    A_{h\to ggg}^{(0)\lambda}(s_{12},s_{23},s_{13}) &= c_{ggh}A_{h,ggg}^{(0)}+\sum_qc_qA_{q,ggg}^{(0)\lambda}(s_{12},s_{23},s_{13})
\end{align}
with
\begin{align}
    A_{h,ggg}^{(0)} &= 1.
\end{align}
Out of the eight possible helicity configurations, only two are independent, the others being related by parity and relabelings. They are given by
\begin{align}
  \mathcal{H}^{+++}(s_{12},s_{23},s_{13})
  &= \mathcal{H}^{---}(s_{12},s_{23},s_{13})
    = \frac{m_h^8}{s_{12}s_{23}s_{13}}\,,\\
  \mathcal{H}^{++-}(s_{12},s_{23},s_{13})
  &= \mathcal{H}^{--+}(s_{12},s_{23},s_{13})
        = \mathcal{H}^{+-+}(s_{23},s_{13},s_{12})\notag\\
        = \mathcal{H}^{-+-}(s_{23},s_{13},s_{12})
        &= \mathcal{H}^{-++}(s_{13},s_{12},s_{23})
  = \mathcal{H}^{+--}(s_{13},s_{12},s_{23})
    = \frac{s_{12}^3}{s_{23}s_{13}}\,.
\end{align}
The quark-loop functions $A_{q,ggg}^{(0)\lambda}$ read
\begin{align}
  A_{q,ggg}^{(0)+++}(s_{12},s_{23},s_{13}) &= A_{q,ggg}^{(0)---}(s_{12},s_{23},s_{13})\,,
\label{eq:Aqggg_hel_relationsa}
\end{align}
 \begin{align}                                              
   A_{q,ggg}^{(0)++-}(s_{12},s_{23},s_{13}) &= A_{q,ggg}^{(0)--+}(s_{12},s_{23},s_{13})
        = A_{q,ggg}^{(0)+-+}(s_{23},s_{13},s_{12})\notag\\
        = A_{q,ggg}^{(0)-+-}(s_{23},s_{13},s_{12})
        &= A_{q,ggg}^{(0)-++}(s_{13},s_{12},s_{23})
        = A_{q,ggg}^{(0)+--}(s_{13},s_{12},s_{23})\label{eq:Aqggg_hel_relationsb}
\end{align}
and can be found in the literature, see Table~\ref{tab:hggrgggrefs}.

\begin{table}[t]
    \centering
    \renewcommand{\arraystretch}{2}
    \begin{tabular}{l|c|c}
            refs. &  $A_{q,ggg}^{(0)+++}(s_{12},s_{23},s_{13})$ & $A_{q,ggg}^{(0)++-}(s_{12},s_{23},s_{13})$ \\
        \hline
            \cite{Ellis:1987xu} & $-2A_4(s_{12},s_{23},s_{13})$ & $\displaystyle-2\frac{m_h^4}{s_{12}^2}A_2(s_{12},s_{23},s_{13})$ \\[10pt]
            \cite{Baur:1989cm} & $\displaystyle\frac{s_{12}s_{23}s_{13}}{m_h^4}\frac{m_f^2}{16}\frac{\M_{++-}}{m_f^2\Delta}$ & $\displaystyle-\frac{s_{23}s_{13}}{s_{12}}\frac{m_f^2}{16}\frac{\M_{+++}}{m_f^2\Delta}$ \\[10pt]
            \cite{Rozowsky:1997dm} & $\displaystyle-i(4\pi)^2v\frac{\zA{12}\zA{23}\zA{31}}{m_h^4}A_4^f(1^+,2^+,3^+,H)$ & $\displaystyle i(4\pi)^2v\frac{\zB{23}\zB{31}}{\zB{12}^3}A_4^f(1^+,2^+,3^-,H)$
    \end{tabular}
    \caption{References for the two independent helicity amplitudes for the quark-loop contribution to the $ggg$ channel, including the relations of their expressions to our notation. In~\cite{Baur:1989cm}, $\Delta=\sqrt{s_{12}s_{23}s_{13}/8}$ is defined and a sign flip for the parity transformed amplitude $\mathcal{M}_{-\lambda}$ is introduced, which is not reflected by (\ref{eq:Aqggg_hel_relationsa}) and (\ref{eq:Aqggg_hel_relationsb}), and instead absorbed into the square-root of $\mathcal{H}^\lambda$. It drops out after squaring the amplitude. In~\cite{Rozowsky:1997dm}, the spinor-helicity formalism~\cite{DeCausmaecker:1981jtq,Kleiss:1985yh,Xu:1986xb,Dixon:2013uaa,Elvang:2013cua} is employed using the convention $\zA{ij}\zB{ji}=s_{ij}$.}
    \label{tab:hggrgggrefs}
\end{table}

\subsubsection[$h\to gq\qb$ channel]{\boldmath $h\to gq\qb$ channel\unboldmath}
\label{subsubsec:Rgqqb}
Denoting the outgoing gluon, quark and anti-quark momenta by $p_1$, $p_q$ and  $p_{\qb}$, respectively, and defining the Mandelstam variables $s_{ij}$ as before with $s_{q\qb}+s_{1q}+s_{1\qb}=m_h^2$, the contribution of the $gq\qb$ channel, for a single massless quark flavour, to the NLO decay rate is given by
\begin{align}
    \Gamma_{h\to gg}^{R,gq\qb} &= \frac{\alpha_s^3}{4\pi m_hv^2}(N_c^2-1)\int\mathrm{d}\Phi_3(p_1,p_q,p_{\qb})\frac{s_{1q}^2+s_{1\qb}^2}{s_{q\qb}}\abs*{A_{h\to gq\qb}^{(0)}(s_{q\qb},s_{1q},s_{1\qb})}^2\,.\label{eq:hggRgqqb}
\end{align}
We already carried out the helicity sum, as the function
\begin{align}
    A_{h\to gq\qb}^{(0)}(s_{q\qb},s_{1q},s_{1\qb}) = c_{ggh}A_{h,gq\qb}^{(0)}+\sum_qc_qA_{q,gq\qb}^{(0)}(s_{q\qb},s_{1q},s_{1\qb})
\end{align}
is helicity independent. The normalisation is again chosen such that for the tree-level contribution
\begin{align}
   A_{h,gq\qb}^{(0)} = 1\,.
\end{align}
The relevant expressions for the quark-loop function can again be found in the literature, see Table~\ref{tab:hggrgqqbrefs}.

\begin{table}[t]
    \centering
    \renewcommand{\arraystretch}{2}
    \begin{tabular}{l|c}
      refs. & $A_{q,gq\qb}^{(0)}(s_{q\qb},s_{1q},s_{1\qb})$ \\
      \hline
      \cite{Ellis:1987xu} & $\displaystyle\frac{m_h^2}{m_h^2-s_{q\qb}}A_5(s_{q\qb},s_{1q},s_{1\qb})$ \\
      \cite{Baur:1989cm} & $\displaystyle\frac{2}{m_h^2-s_{q\qb}}\mathcal{A}(s_{q\qb},s_{1q},s_{1\qb})$
    \end{tabular}
  \caption{References for the quark-loop amplitudes contributing to the
    $gq\qb$ channel and relations to the corresponding expressions therein.}
    \label{tab:hggrgqqbrefs}
\end{table}

\subsection{Implementation and validation}
\label{subsec:code}
We implemented the calculation of the $h\to gg$ decay rate as a C++ program. For the real radiation phase space integration we used the Monte Carlo algorithm SUAVE, implemented in the CUBA library~\cite{Hahn:2004fe}. Infrared phase space singularities and the cancellation of infrared poles of the virtual correction matrix elements are handled by means of the antenna subtraction method~\cite{GehrmannDeRidder:2005cm,Daleo:2006xa,Currie:2013vh}, see Appendix~\ref{app:IRsub}. We use the CRunDec package~\cite{Chetyrkin:2000yt,Schmidt:2012az,Herren:2017osy} to obtain the numerical value of the running strong coupling in the $N_F=5$ scheme at two-loop level. If required, the same package can also be used to convert between the different quark mass schemes of eqs.~(\ref{eq:Xscheme1}) and~(\ref{eq:Xscheme2}).

We validated our implementation by checking the individual ingredients and stages of the calculation:

\paragraph{Validation of the amplitudes:}
We have recalculated all amplitudes except the two-loop NLO virtual SM contributions. Our analytic formulae agree with the ones given in the literature. Furthermore, after recovering the ordinary SM amplitudes by setting the effective couplings to their appropriate values, $c_{ggh}=0$ and $c_t=c_b=1$, we compared the squared one-loop Born and real correction amplitudes to their numerical counterparts obtained from OpenLoops2~\cite{Buccioni:2019sur}, evaluated at a set of random phase space points. We found agreement within machine precision. OpenLoops2 also provides the relevant tree level and one-loop amplitudes for $h\to gg$ at NLO QCD in the heavy-top limit, so that we could validate all amplitudes with effective $hgg$ vertex setting $c_t=c_b=0$ and $c_{ggh}=\frac{2}{3}\left(1+11\frac{\alpha_s}{4\pi}\right)$~\cite{Spira:1995rr}\footnote{\label{ftn:cggh_ht}Note that expanding $c_{ggh}$ in terms of the strong coupling interferes with the perturbative expansion of the $h\to gg$ amplitude. The only place where the additional term $11\frac{\alpha_s}{4\pi}$ is relevant at $\order{\alpha_s^3}$ is as a finite renormalisation of the virtual corrections, proportional to the Born amplitude
(see also Appendix \ref{app:NNLOfix}).}.
We are not aware of any numerical implementation of the two-loop amplitude,
against which we could compare in the same way. The references displayed in Table \ref{tab:hggVrefs} are in fact numerically self-consistent. However, only \cite{Anastasiou:2006hc} contains the unrenormalized amplitudes, for which we have explicitly carried out the renormalization procedure.

\paragraph{Validation of the infrared subtraction:}
        Due to the simple infrared structure of the process and the fact that the decay $h\to gg$ in the limit of infinite top mass is actually used to derive the gluon-gluon antenna functions~\cite{GehrmannDeRidder:2005aw}, we can check by hand that the infrared poles of the virtual correction amplitudes are correctly cancelled.  We confirmed the cancellation numerically as a test of our implementation. In order to validate the proper functioning of the real subtraction, we checked numerically that the ratio between real radiation matrix elements and the subtraction term tends to unity as we probe regions of the phase space ever closer to IR singular configurations. For the actual phase space integration, we implemented a cut parameter preventing the integrator to probe regions close to the singularity where cancellations between a very large matrix element and subtraction term occur. Those cancellations can escape the numerical precision leading to instabilities. The value of this cut must be chosen small enough so that only regions of the phase space are cut out in which the matrix element and the subtraction term can be treated as equal, leading to a vanishing contribution to the full integral. The final result should then not depend on small variations of the cut parameter. We checked that this is indeed the case.
        
\paragraph{Validation of the decay rate:}
The code is structured such that we can vary the effective couplings at will. By setting them to the appropriate values, we can therefore calculate the $h\to gg$ decay rate up to NLO QCD either with full $m_t$-dependence or in the limit of infinite top mass\footnote{The latter without actually sending $m_t\to\infty$, but rather by setting $c_{ggh}$ to the value of the effective coupling in the heavy-top limit, see footnote~\ref{ftn:cggh_ht}.}. We checked that for both cases we can reproduce the numerical results known from the literature~\cite{Spira:1995rr} within uncertainties. As a consistency check we confirmed that when setting $c_{ggh}$ to the heavy-top limit value, $c_t$ to $1$ and sending $m_t\to\infty$, the rate tends to four times the rate in the
heavy-top limit, as expected. This tests the relative phase between the $c_{ggh}$ term and the quark-loop amplitudes. The $h\to gg$ rate including a local $hgg$ vertex has been implemented previously in the program eHDECAY~\cite{Contino:2014aaa} considering QCD effects up to N$^3$LO in the limit of heavy fermion masses. The exact dependence on the top and bottom mass is included up to NLO in the pure fermion loop contributions, corresponding to our coefficients $A_{tt}$, $A_{bb}$ and $A_{bt}$. eHDECAY also takes the charm quark into account, which we neglected in our numerical studies due to its small overall impact. In order to compare with eHDECAY we included it as well. Checking different settings for the effective couplings $c_i$ we observe between $5\%$ and $12\%$ smaller rates obtained with our code, which can be attributed to the missing higher order QCD effects in our implementation. This discrepancy, however, is fully covered by the NLO scale uncertainty (see section~\ref{sec:pheno}). Using additional information on the higher order effects in the heavy fermion mass limit from~\cite{Kramer:1996iq,Chetyrkin:1997un,Schroder:2005hy,Chetyrkin:2005ia,Baikov:2006ch,Spira:2016zna}, we can supplement the results from our code with those effects, obtaining the rate at the same order in QCD as eHDECAY. Doing so we find agreement with eHDECAY for all tested configurations of effective couplings.


\section{\boldmath Phenomenological results for $h\to gg$ \unboldmath}
\label{sec:pheno}

The coefficients $A_i$ introduced in eqs.~(\ref{eq:rateLO}) and (\ref{eq:rateNLO}) can be obtained by calculating the rate for six different combinations of coupling values and solving a simple system of linear equations, yielding
\begin{align}
    A_{gg} &= \Gamma_{h\to gg}\Big|_{c_{ggh}=1,c_t=0,c_b=0}\,,\\
    A_{tt} &= \Gamma_{h\to gg}\Big|_{c_{ggh}=0,c_t=1,c_b=0}\,,\\
    A_{bb} &= \Gamma_{h\to gg}\Big|_{c_{ggh}=0,c_t=0,c_b=1}\,,\\
    A_{tg} &= \Gamma_{h\to gg}\Big|_{c_{ggh}=1,c_t=1,c_b=0} - A_{gg} - A_{tt}\,,\\
    A_{bg} &= \Gamma_{h\to gg}\Big|_{c_{ggh}=1,c_t=0,c_b=1} - A_{gg} - A_{bb}\,,\\
    A_{bt} &= \Gamma_{h\to gg}\Big|_{c_{ggh}=0,c_t=1,c_b=1} - A_{tt} - A_{bb}\,.
\end{align}
We compute the coefficients using the input parameters shown in Table~\ref{tab:input}, treating the charm quark as massless, i.e. neglecting the decay of the Higgs through charm quark loops. In the SM this contribution accounts for less than $3\%$ of the rate.

\begin{table}[t]
\centering
    \begin{tabular}{c|cp{1.2cm}p{1.2cm}|cp{1.2cm}p{1.2cm}}
        & \multicolumn{3}{c|}{LO} & \multicolumn{3}{c}{NLO} \\
        coefficient & value $[\MeV]$ & param. uncert. & scale uncert. & value $[\MeV]$ & param. uncert. & scale uncert. \\
        \hline
        $A_{gg}$ & $0.41360$ & $\pm1.5\%$ & ${}^{+23\%}_{-17\%}$ & $0.59755$ & $\pm1.7\%$ & ${}^{+9.5\%}_{-9.6\%}$ \\[5pt]
        $A_{tt}$ & $0.19595$ & $\pm1.5\%$ & ${}^{+23\%}_{-17\%}$ & $0.32290$ & $\pm1.8\%$ & ${}^{+13\%}_{-11\%}$ \\[5pt]
        $A_{tt}+\delta A_{tt}$ & & --- & & $0.32468$ & $\pm1.8\%$ & ${}^{+13\%}_{-12\%}$ \\[5pt]
        $A_{bb}$ & $0.00218$ & $\pm4.0\%$ & ${}^{+23\%}_{-17\%}$ & $0.00328$ & $\pm4.0\%$ & ${}^{+11\%}_{-10\%}$ \\[5pt]
        $A_{tg}$ & $0.56937$ & $\pm1.5\%$ & ${}^{+23\%}_{-17\%}$ & $0.88041$ & $\pm1.8\%$ & ${}^{+11\%}_{-11\%}$ \\[5pt]
        $A_{bg}$ & $-0.03442$ & $\pm2.0\%$ & ${}^{+23\%}_{-17\%}$ & $-0.04837$ & $\pm2.2\%$ & ${}^{+8.8\%}_{-9.2\%}$ \\[5pt]
        $A_{bt}$ & $-0.02369$ & $\pm2.0\%$ & ${}^{+23\%}_{-17\%}$ & $-0.03569$ & $\pm2.2\%$ & ${}^{+11\%}_{-10\%}$
    \end{tabular}
\caption{Values of the LO and NLO coefficients. The parametric uncertainty is derived by varying the input parameters, the scale uncertainty by varying the renormalisation scale $\mu_R$ by factors of $0.5$ and $2$.}
\label{tab:coeffs}
\end{table}

Table~\ref{tab:coeffs} shows the results for LO and NLO QCD together with the value of the coefficient $A_{tt}^{NLO}$ shifted by
\begin{align}
    \delta A_{tt}^{NLO} &= \left(\frac{\alpha_s}{4\pi}\right)^4\frac{242}{9\pi}\frac{m_h^3}{v^2}\qquad(\text{for }N_c=3)\,.\label{eq:NNLOfix}
\end{align}
Formally this contribution is part of the NNLO corrections. It should, however, not be neglected, because it is effectively the dominant contribution to the rate close to specific values of the couplings $c_{ggh}$, $c_t$ and $c_b$ for which the LO amplitude is parametrically suppressed, but the higher order corrections are not. In these regions of the parameter space the rate truncated at $\order{\alpha_s^3}$ can even become negative and thus unphysical. It is important to distinguish these configurations from a parametric suppression where simply all couplings are chosen to be small, which would affect the rate at all orders in a similar way.

We will include the shift in all our NLO plots in order to get more reliable predictions in these regions of the parameter space. One should keep in mind that these predictions are effectively of lower order in perturbation theory and are subject to larger uncertainties compared to the rate away from any parametric suppression, where the usual perturbative expansion holds and the phenomenological impact of the shift is small. In Appendix~\ref{app:NNLOfix}, we motivate and derive equation~(\ref{eq:NNLOfix}) in more detail.

Besides the numerical values of the coefficients, we also give their respective scale and parametric uncertainties\footnote{The Monte Carlo error from the numerical phase space integration is several orders of magnitude smaller than those uncertainties.}. The former is obtained from varying the renormalization scale $\mu_R$ by factors of $0.5$ and $2$ around its central value, which we choose to be $\mu_R=m_h$. The scale uncertainty serves as an estimate of the impact of missing higher order corrections, as the all-order result must be scale independent.

The parametric uncertainty is derived by varying the input parameters within their respective errors. Treating them as uncorrelated, we change one at a time while keeping all others fixed and eventually sum the individual variations, which are approximately symmetric, in quadrature. It is important to notice that the resulting parametric uncertainties on the coefficients $A_i$ are correlated. The corresponding correlation matrix is given in Appendix~\ref{app:correlations}. For all coefficients but $A_{bb}$ the uncertainty on the value of the strong coupling $\alpha_s(m_Z)$ has the largest impact. The uncertainty on $A_{bb}$ is driven mainly by that on $m_b$, which also has small impact on $A_{bg}$ and $A_{bt}$. The uncertainties of the other input parameters are negligible in comparison.

Using the results for the coefficients in Table~\ref{tab:coeffs} and eqs.~(\ref{eq:rateLO}) and (\ref{eq:rateNLO}) for the LO and NLO decay rates, respectively, we can now determine the rate for arbitrary values of the effective couplings.
For the SM case ($c_{ggh}=0$, $c_t=c_b=1$) we find
\begin{align}
    \Gamma_{h\to gg}^{LO,SM} &= \left(0.1744\pm1.5\%(\text{param.})^{+23\%}_{-17\%}(\text{scale})\right)\MeV\,,\label{eq:rateSMLO}\\[5pt]
     \Gamma_{h\to gg}^{NLO,SM} &= \left(0.2923\pm1.8\%(\text{param.})^{+13\%}_{-12\%}(\text{scale})\right)\MeV\,.\label{eq:rateSMNLO}
\end{align}
The contribution of the shift~(\ref{eq:NNLOfix}) to the NLO rate is less than $1\%$. It is included in (\ref{eq:rateSMNLO}).

The contribution of the bottom-quark loops, entering through $A_{bb}$, $A_{bt}$ and $A_{bg}$ is small, but non-negligible compared to the largest SM contribution $A_{tt}$, the top-quark loop. Indeed, the bottom loop interferes destructively with the top loop and decreases the decay rate by about $10\%$ compared to the top-only case, both at LO and NLO.

The coefficients related to the local Higgs-gluon interaction, $A_{gg}$, $A_{tg}$ and $A_{bg}$, are comparatively large. In particular we observe a strong interference between the local and top loop contribution. This renders the rate very sensitive to not only the effective coupling $c_{ggh}$, but also the ratio $c_{ggh}/c_t$.

\begin{figure}
\centering
    \begin{subfigure}[b]{0.45\textwidth}
    \centering
        \includegraphics[width=\textwidth]{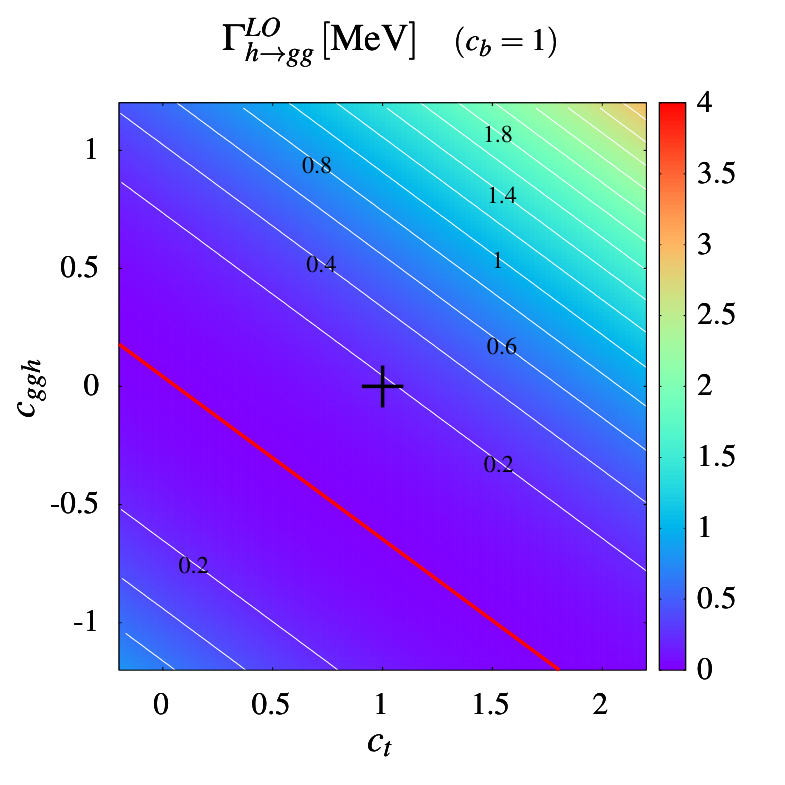}
    \end{subfigure}
    \begin{subfigure}[b]{0.45\textwidth}
    \centering
        \includegraphics[width=\textwidth]{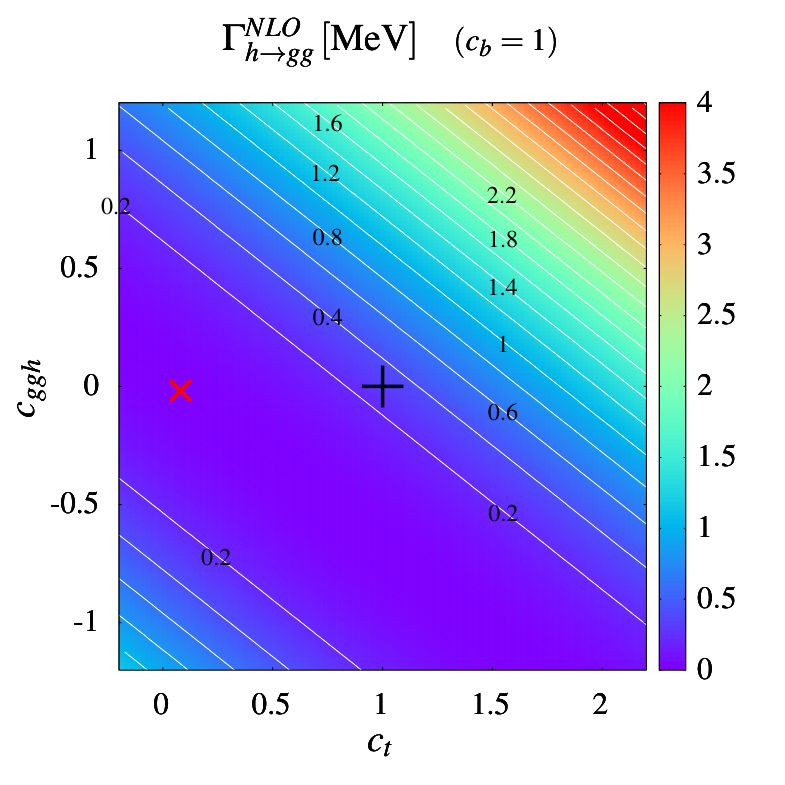}
    \end{subfigure}
    \begin{subfigure}[b]{0.45\textwidth}
    \centering
        \includegraphics[width=\textwidth]{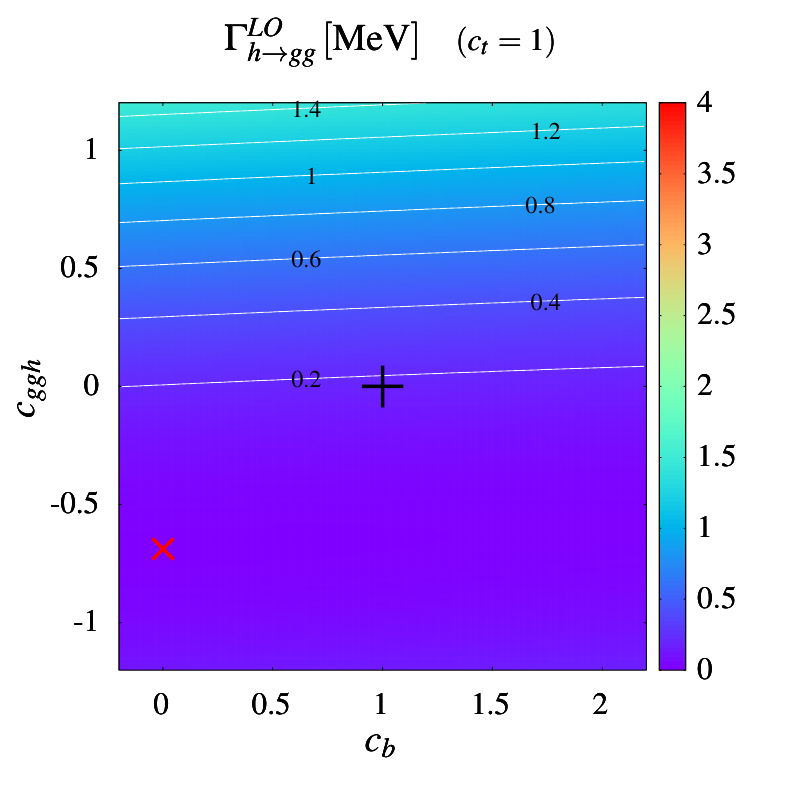}
    \end{subfigure}
    \begin{subfigure}[b]{0.45\textwidth}
    \centering
        \includegraphics[width=\textwidth]{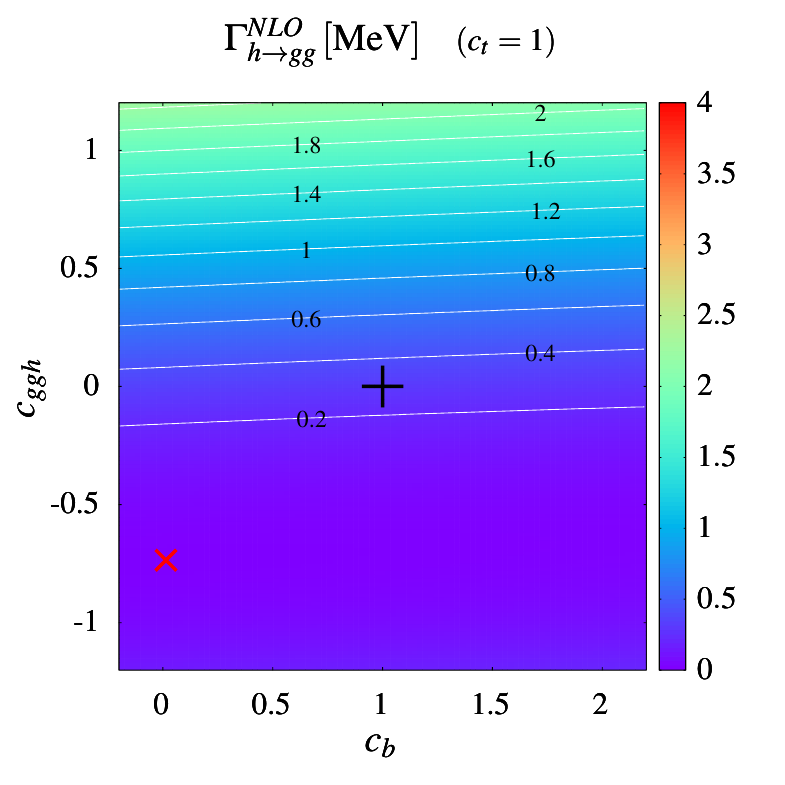}
    \end{subfigure}
    \begin{subfigure}[b]{0.45\textwidth}
    \centering
        \includegraphics[width=\textwidth]{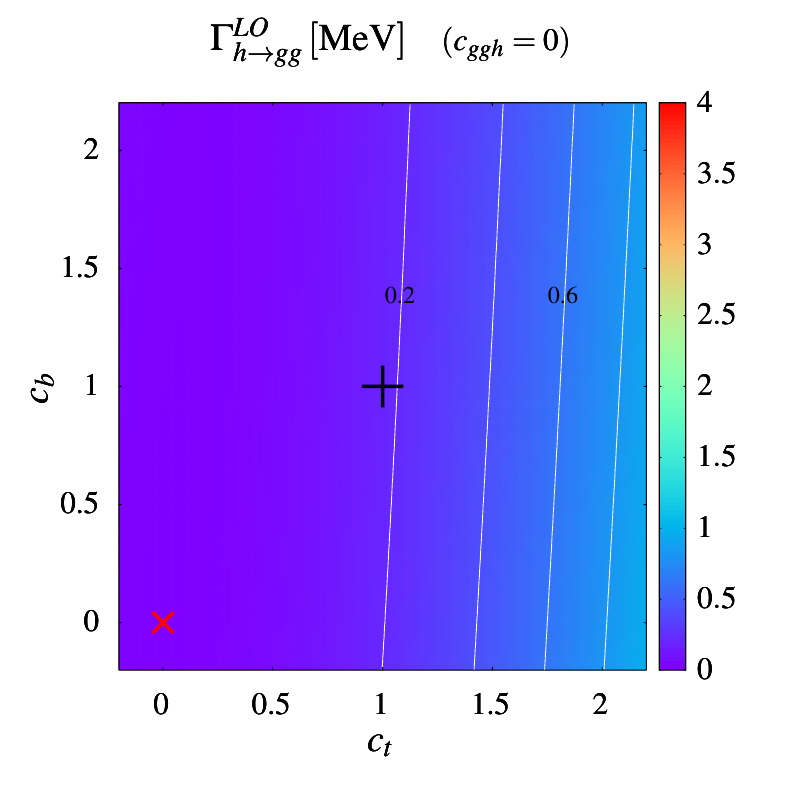}
    \end{subfigure}
    \begin{subfigure}[b]{0.45\textwidth}
    \centering
        \includegraphics[width=\textwidth]{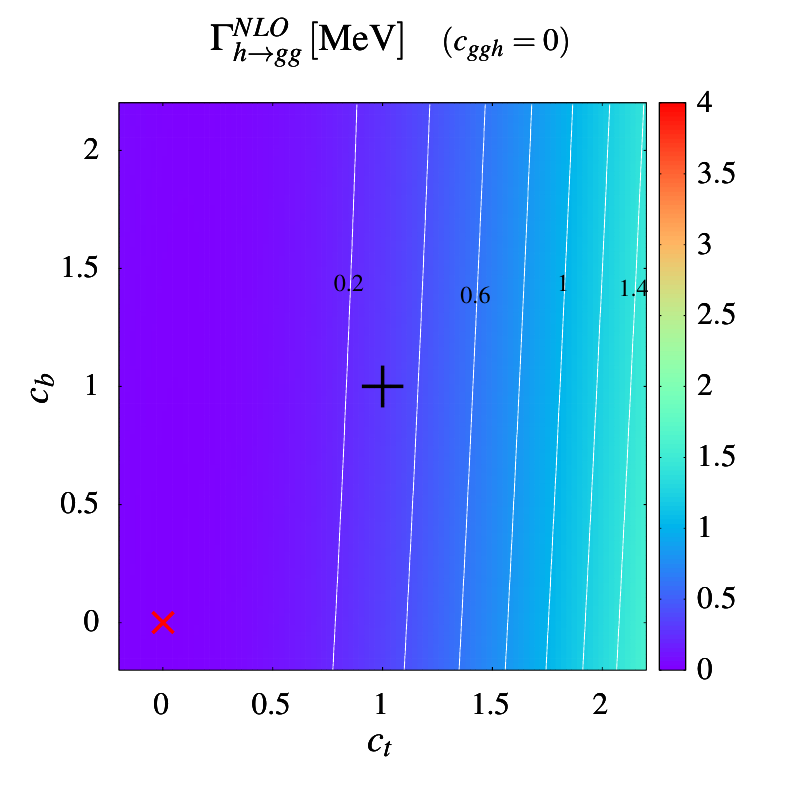}
    \end{subfigure}
    \caption{Contour plots of the $h\to gg$ LO (left) and NLO (right) decay rates for different values of the effective couplings. The SM configuration is marked with a black cross at the centre of the plots. A red cross shows the global minimum. In the upper left panel, LO for fixed $c_b=1$, the global minimum is not a single point but rather a line.}
    \label{fig:contour}
\end{figure}

\begin{figure}
\centering
    \includegraphics[width=0.6\textwidth]{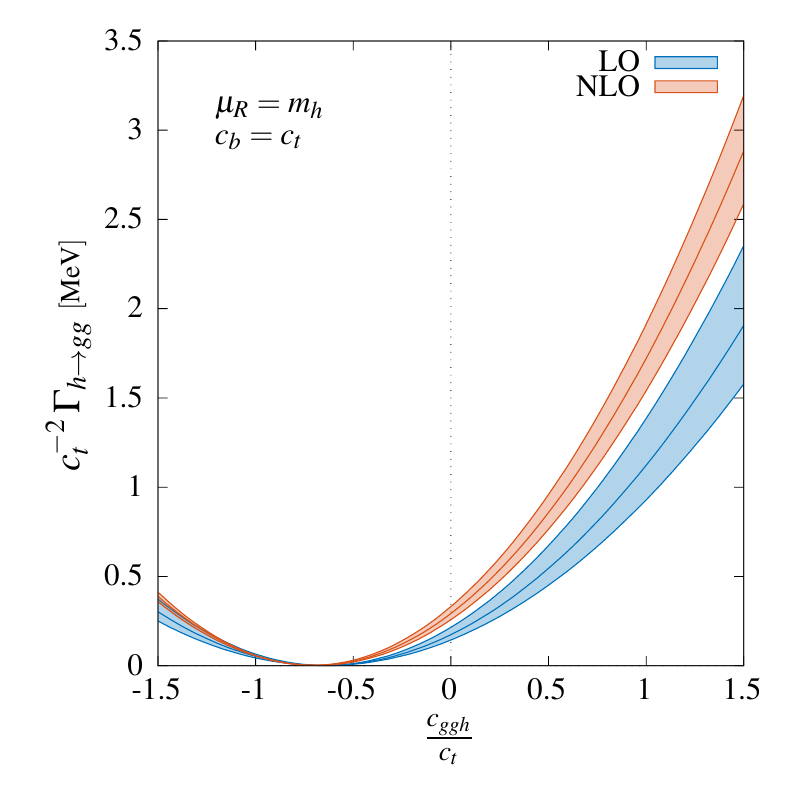}
    \caption{Inclusive decay rate $\Gamma_{h\to gg}$, rescaled by $c_t^{-2}$, as a function of the ratio $c_{ggh}/c_t$ of the effective couplings. The error bands are obtained through scale variation.}
    \label{fig:bands}
\end{figure}

We can confirm these findings graphically by plotting heat maps of the decay rate, varying two of the effective couplings around their SM values, while keeping the third fixed at its SM value. Figure~\ref{fig:contour} shows the LO (left panels) and NLO (right panels) rates. The SM configuration $c_{ggh}=0$ , $c_t=c_b=1$ always lies at the centre and is marked with a black cross. In addition we highlighted the global minimum with a red cross. At LO it is given by
\begin{align}
    \evalu{\Gamma_{h\to gg}^{LO}}{\mathrm{min}} &= c_b^2\left(\frac{\alpha_s}{4\pi}\right)^2\frac{N_c^2-1}{64\pi}\frac{m_h^3}{v^2}\left(\mathrm{Im}\left\{A_q^{(0)}(\tau_b)\right\}\right)^2\,,
\end{align}
and is located where the real part of the LO amplitude~(\ref{eq:Ahgg0}) vanishes, i.e.
\begin{align}
    c_{ggh}A_h^{(0)}+\frac{1}{2}c_tA_q^{(0)}(\tau_t)+\frac{1}{2}c_b\mathrm{Re}\left\{A_q^{(0)}(\tau_b)\right\} &= 0\,.
\end{align}
Here we made use of the fact that $A_h^{(0)}$ and $A_q^{(0)}(\tau_t)$ (eqs.~(\ref{eq:Ah0}) and~(\ref{eq:Al0Aq0})) are real and only  $A_q^{(0)}(\tau_b)$ has an imaginary part. Consequently, the global minimum in the upper left panel, LO with $c_b=1$ fixed, is not a single point, but rather a line. Close to the minimum, both the LO and NLO rates are much smaller than the rates for the SM like configuration, (\ref{eq:rateSMLO}) and~(\ref{eq:rateSMNLO}), due to the parametric suppression discussed above. This can also be seen in Figure~\ref{fig:bands}, where the LO and NLO decay rates, normalized to $c_t^{-2}$, as a function of the ratio $c_{ggh}/c_t$ of the effective couplings are shown. The scale uncertainty reduces when going from LO to NLO, but the scale bands do not overlap, hinting towards a slow convergence of the perturbative series. This suggests that the calculation of NNLO corrections might be needed to get a more reliable theory estimate of the decay rate. We remark that in the region of parametric suppression of the rates, the scale variation is not a good measure for the theory uncertainty: the underlying assumption that consecutive terms in the perturbative expansion decrease in magnitude is not fulfilled.

\begin{figure}
\centering
    \begin{subfigure}[b]{0.45\textwidth}
    \centering
        \includegraphics[width=\textwidth]{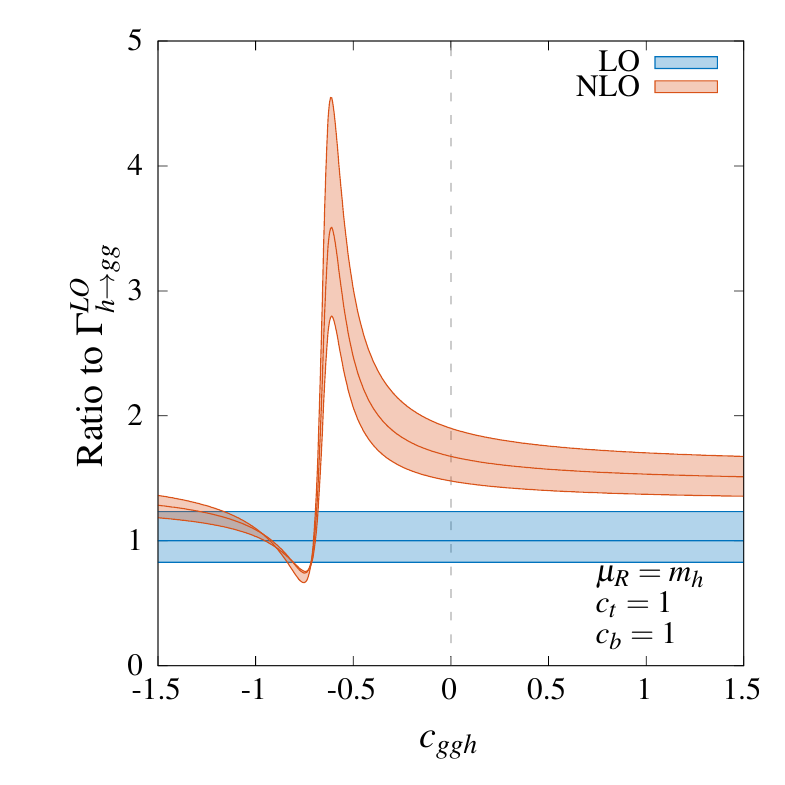}
        \caption{}
        \label{fig:Kfactor_ggh}
    \end{subfigure}  
    \begin{subfigure}[b]{0.45\textwidth}
    \centering
        \includegraphics[width=\textwidth]{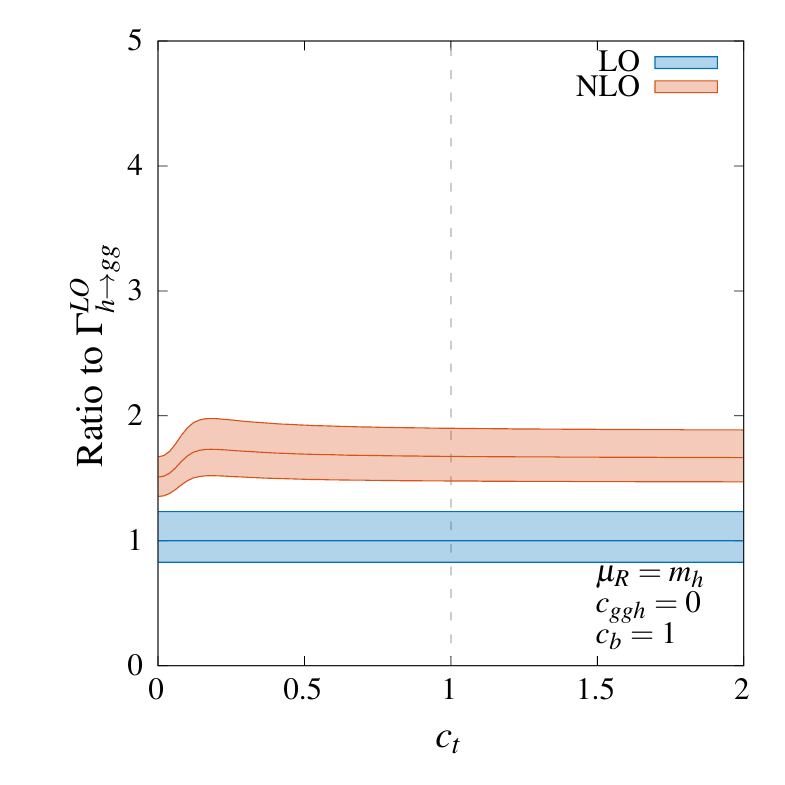}
        \caption{}
        \label{fig:Kfactor_t}
    \end{subfigure}
    \begin{subfigure}[b]{0.45\textwidth}
    \centering
        \includegraphics[width=\textwidth]{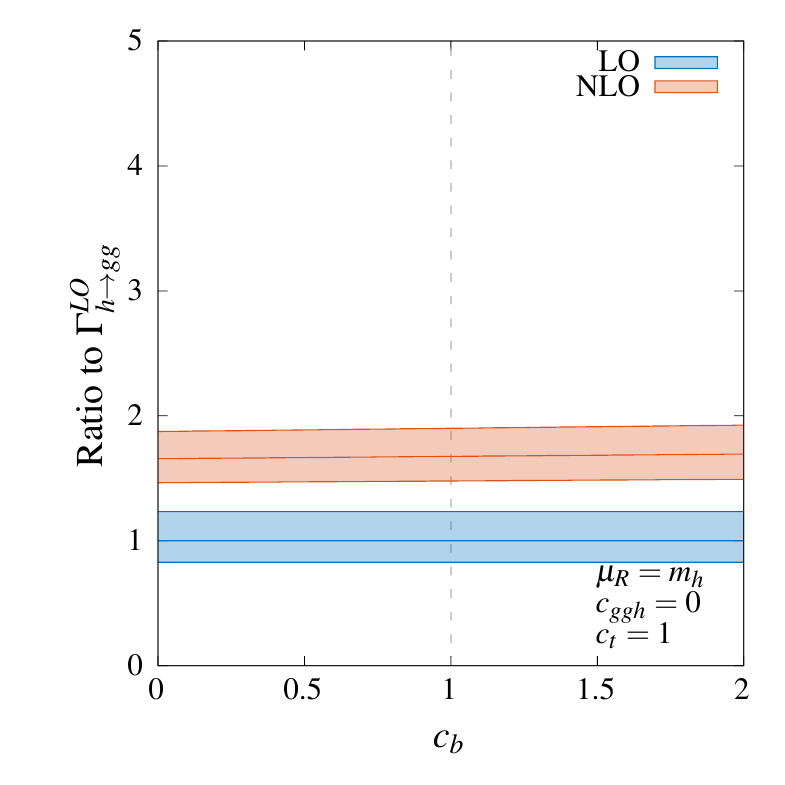}
        \caption{}
        \label{fig:Kfactor_b}
    \end{subfigure}
    \caption{Dependence of the QCD $K$-factor on the values of the effective couplings. In each panel one coupling is varied while the other two are kept at their respective SM values. Again, the error bands are obtained through scale variation. The denominator $\Gamma_{h\to gg}^{LO}$ is fixed to the central scale $\mu_R=m_h$.}
    \label{fig:Kfactor}
\end{figure}

The effect of the NLO corrections can also be studied by plotting the QCD K-factor, i.e. the ratio $\Gamma_{h\to gg}^{NLO}/\Gamma_{h\to gg}^{LO}$, as a function of the effective couplings $c_{ggh}$, $c_t$ and $c_b$. In Figure~\ref{fig:Kfactor} we show the K-factor, varying one coupling at a time and fixing the others at their respective SM values. The denominator is evaluated at the central scale $\mu_R=m_h$, while in the numerator $\mu_R$ is varied as previously by factors of $0.5$ and $2$. We also show the LO in blue, so that we can easily compare the scale uncertainty to the magnitude of the shift induced by the QCD corrections. The SM like configuration is indicated with a dashed line. We again see that the bands mostly do not overlap, pointing towards the necessity of higher order corrections to be included.

Figure~\ref{fig:Kfactor_ggh} shows the K-factor for $c_t=c_b=1$ as a function of $c_{ggh}$. For $c_{ggh}>0$, the QCD corrections increase the rate by about $60\%$, with only a modest dependence on $c_{ggh}$, whereas for $c_{ggh}<0$, in particular close to $c_{ggh}\approx-0.7$, the K-factor shows a highly non-trivial behaviour. Here the effect of the parametric suppression of the LO rate can be seen. As explained above, the scale band of the NLO result underestimates the true uncertainty at this point.

Figure~\ref{fig:Kfactor_t} captures the dependence on $c_t$ for $c_b=1$ and $c_{ggh}=0$. The K-factor is almost flat, showing an increment of the rate of about $70\%$. Close to $c_t=0$ the K-factor slightly drops. Here the destructive interference between the top and bottom loop is enhanced, with an effect similar to what we saw in the previous plot, just less pronounced.

The last plot, Figure~\ref{fig:Kfactor_b}, shows the K-factor as a function of $c_b$, for $c_t=1$ and $c_{ggh}=0$. In accordance with the small overall contribution from the bottom-quark amplitude, we see a negligible dependence of the K-factor on $c_b$ in this parameter range\footnote{We see a slightly increased dependence if we enlarge the parameter range of $c_b$ to $\pm\order{10}$.} Going from LO to NLO increases the rate by approximately $70\%$.

In the figures we discussed we allow $\order{1}$ deviations of the effective couplings $c_{ggh}$, $c_t$ and $c_b$ from their respective SM values. A global fit~\cite{deBlas:2018tjm} of the parameters of the EWChL to Higgs-boson signal strength measurements at the LHC shows that only deviations of $\order{0.1}$ are allowed by data, neglecting a few configurations of the couplings which are deemed unnatural in the context of the EFT approach. The fit in~\cite{deBlas:2018tjm} has been performed at LO in QCD, with the exception of approximate NLO effects in the top quark contribution to the $h\to\gamma\gamma$ and $h\to gg$ decay rates, in the $m_t\to \infty$ limit. With our calculation we provide one ingredient to extend such a fit to include full NLO QCD effects.


\section{\boldmath Results for $h\to gg$ and $h\to \gamma\gamma$ in SMEFT\unboldmath}
\label{sec:smeft}

\begin{table}
\centering
\begin{tabular}{c|r|c}
    Coupling & EWChL & Warsaw basis\\
    \hline
    $c_{f}$ & $\frac{{\cal M}_{f,1}}{m_f}$ &$1+\frac{v^2}{\Lambda^2}C_{\varphi\Box}-\frac{v^2}{4\Lambda^2}C_{\varphi D}-\frac{v^3}{\sqrt{2}m_f\Lambda^2}C_{f\varphi}$ \\
    $c_V$ & $\frac{F_1}{2}$ & $1+\frac{v^2}{\Lambda^2}C_{\varphi\Box}-\frac{v^2}{4\Lambda^2}C_{\varphi D}$\\
    $c_{ggh}$ & $16\pi^2f_{Xh3,1}$ &$\frac{32\pi^2 v^2}{g_s^2\Lambda^2}C_{\varphi G}$\\
    $c_{\gamma\gamma h}$ & $16\pi^2\left(2f_{Xh1,1}+f_{Xh2,1}+f_{XU1,1}\right)$ & $\frac{32\pi^2 v^2}{\Lambda^2}\left(\frac{C_{\varphi W}}{g^2}+\frac{C_{\varphi B}}{g'^2}-\frac{C_{\varphi WB}}{gg'}\right)$\\
\end{tabular}
\caption{Definition of the anomalous couplings $c_i$ in (\ref{eq:ewchl}) in terms of the fundamental parameters of the EWChL defined in Section \ref{sec:intro} and reference \cite{Buchalla:2013rka} and the Warsaw basis Wilson coefficients $C_i$.}\label{tab:SMEFTEWChL}
\end{table}
The previous discussion was based on the anomalous couplings in the context of the EWChL, which parametrizes the Higgs sector in a non-linear manner and is particularly suited for Higgs-related BSM scenarios with strong coupling dynamics. However, the results presented in this work are actually applicable to a broad variety of situations.

A common extrapolation of the SM into the UV regime consists in adding higher dimensional operators to the dimension-four SM, resulting in the SMEFT. We restrict ourselves to the leading corrections from operators of canonical dimension six. Here, the electroweak symmetry breaking pattern is realized in a linear manner and the new physics (NP) can decouple from the SM allowing for a large mass gap. It is important to notice, however, that even in such general situations, the power counting is not as arbitrary as it seems at first glance. For instance, it is easy to construct explicit weakly coupled UV models that, when matched to dimension-six terms in the Warsaw basis~\cite{Grzadkowski:2010es}, result in a hierarchy among operators of the same mass dimension. The clue lies in keeping track of explicit loop factors $1/16\pi^2$ arising in the full theory that can be hidden in the Wilson coefficients of certain local operators. A systematic power-counting prescription for SMEFT is therefore defined by canonical dimensions supplemented by a loop-counting rule,
allowing us to keep track of the loop expansion, on which perturbative
calculations in quantum field theory are based~\cite{Buchalla:2022vjp}.

For our purpose, the most significant implication is that operators featuring field strength tensors (e.g. the operator\footnote{Here and in the following, we employ the notation of~\cite{Dedes:2017zog}.} $Q_{\varphi G}$) are suppressed
with an extra loop factor when compared to the remaining ones. The formalism of the EWChL in (\ref{eq:ewchl}) already accounts for loop factors in modified vertices involving the Higgs boson and can hence be taken over to SMEFT straightforwardly\footnote{The EWChL can account for strong coupling scenarios in the gauge boson sector pushing the first deviations from the SM, parametrized by $F_1$, formally to the LO. This is not the case for weak coupling scenarios which are conveniently handled by SMEFT.}, see Table \ref{tab:SMEFTEWChL}. The coefficients $c_i$ can naturally be taken as ${\cal O}(1)$ numbers. As a consequence, setting $\Lambda=1\TeV$, we have for instance $C_{\varphi G}\approx 0.08\, c_{ggh}$, which makes the implicit loop factor in the definition of $C_{\varphi G}$ manifest. For further comments about the applicability of this Table, see \cite{Heinrich:2022idm}.

\begin{figure}
\centering
\begin{subfigure}[b]{\textwidth}
    \centering
        \includegraphics[scale=0.7]{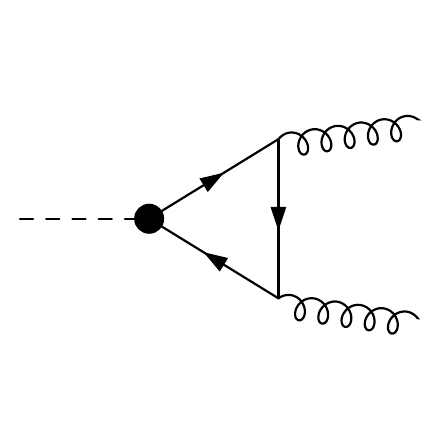}
                ~~~
        \includegraphics[scale=0.7]{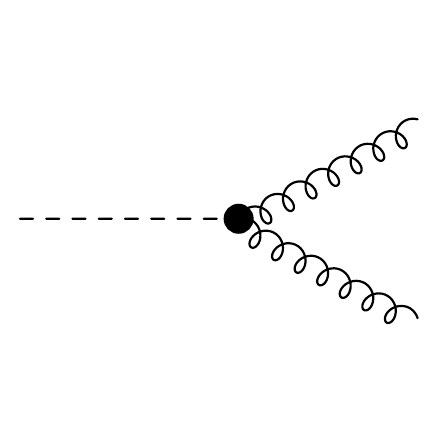}
        \caption{}
    \end{subfigure}

    \vspace{10pt}

    \begin{subfigure}[b]{\textwidth}
    \centering
        \includegraphics[scale=0.7]{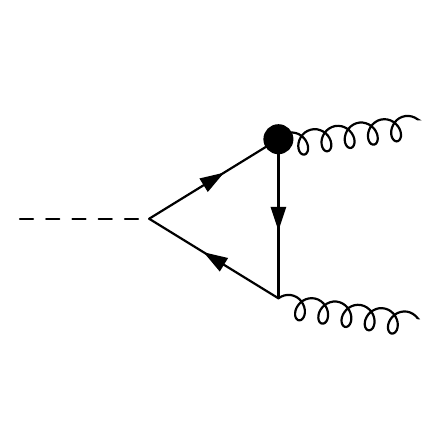}
        ~~~
        \includegraphics[scale=0.7]{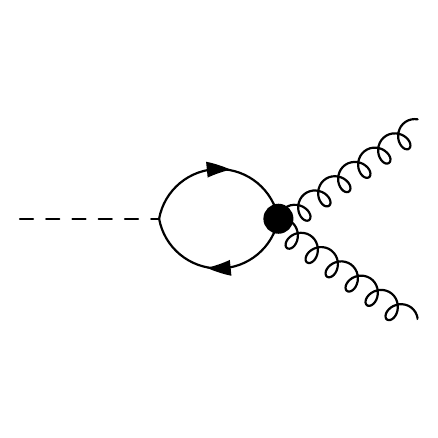}
        ~~~
        \includegraphics[scale=0.7]{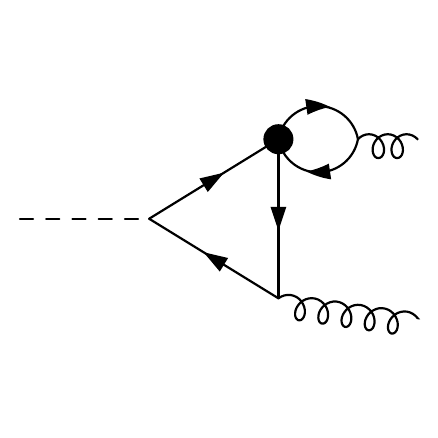}
                ~~~
        \includegraphics[scale=0.7]{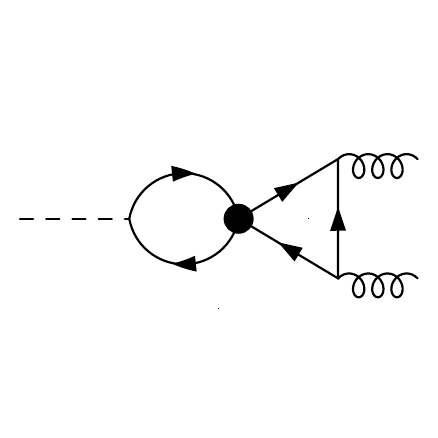}
        \caption{}
    \end{subfigure}
    \caption{Diagrams contributing to $h\to gg$ in SMEFT with single dimension-six insertions (black dots). (a) Contributions of order $\sim g_s^2/16\pi^2$ with anomalous couplings defined in Table \ref{tab:SMEFTEWChL}. (b) Contributions of order $\sim g_s^2/(16\pi^2)^2$ that can consistently be neglected. This qualitative picture can be taken over to the process $h\to\gamma\gamma$.}\label{fig:chromomagnetic}
\end{figure}

In contrast, operators that induce anomalous couplings without the Higgs boson (e.g. the chromomagnetic operator $Q_{u G}$) are present within this framework only at subleading order. They can be neglected consistently without spoiling the underlying systematics. Similar arguments hold true for four-fermion operators. Despite appearing with unsuppressed Wilson coefficients, the relevant diagrams are of an explicit two-loop topology and can thus be dropped, see Figure \ref{fig:chromomagnetic} for $h\to gg$.

Let us emphasize that in the context of NP in the Higgs sector and in particular when considering the Higgs decay channels highlighted in this paper, it is advantageous to work with the EWChL, independently of the actual high energy dynamics being strongly or weakly coupled to the SM. While the difference between the EWChL and SMEFT is less apparent when restricting the latter to canonical dimension six, it becomes more relevant when higher dimensional operators are
considered. For instance, the impact of a generic $(2n+4)$-dimensional operator $(\varphi^\dagger\varphi)^n G_{\mu\nu}^aG^{a\mu\nu}$ to the local Higgs-gluon-gluon interaction in SMEFT is already accounted for by $c_{ggh}$ within the EWChL. While SMEFT has contributions at all orders in the $1/\Lambda$-expansion, only a single coefficient is responsible in the EWChL. An explicit distinction between the various $1/\Lambda^{2n}$-terms that eventually sum up to $c_{ggh}$ is not necessary at this stage as it would increase the number of independent parameters,
complicating the exploration of NP effects, which are yet to be discovered.
Based on the idea of organizing the Higgs-field factors
$(\varphi^\dagger\varphi)^n$ in higher dimensional operators, the framework of geoSMEFT has been developed \cite{Helset:2020yio}, for which an analysis of $h\to gg$ can be found in \cite{Corbett:2021cil,Martin:2021cvs}. \\
In the language of the EWChL and working at NLO in QCD we have
\begin{align}
\frac{\Gamma_{h\to gg}^{\text{EWChL}}}{\Gamma_{h\to gg}^{\text{SM}}}=1+2 \ \delta_{c_t}+2.7116 \ c_{ggh}+\delta_{c_t}^2+1.8404 \ c_{ggh}^2+2.7116 \ \delta_{c_t}c_{ggh}\, , \label{eq:EWChLcorrectionLambda}
\end{align}
where we defined $\delta_{c_t}\equiv c_t-1$. 
Employing the relations of Table \ref{tab:SMEFTEWChL}, it is straightforward to obtain a numerical expression for SMEFT up to operator dimension six and NLO in QCD. Defining $\tilde C_i\equiv C_i v^2/\Lambda^2$, we find
\begin{align}
\frac{\Gamma_{h\to gg}^{\text{SMEFT}}}{\Gamma_{h\to gg}^{\text{SM}}}=1&+2\left(\tilde C_{\varphi\Box}-\frac{1}{4}\tilde C_{\varphi D}\right)-2.0164 \ \tilde C_{t\varphi}+578.04 \ \tilde C_{\varphi G}\nonumber\\
&+\left(\tilde C_{\varphi\Box}-\frac{1}{4}\tilde C_{\varphi D}\right)^2-2.0164 \left(\tilde C_{\varphi\Box}-\frac{1}{4}\tilde C_{\varphi D}\right)\tilde C_{t\varphi}\nonumber\\
&+1.0164 \ \tilde C_{t\varphi}^2+8.3632\cdot 10^{4} \ \tilde C_{\varphi G}^2\nonumber\\
&+578.04\left(\tilde C_{\varphi\Box}-\frac{1}{4}\tilde C_{\varphi D}\right)\tilde C_{\varphi G}-582.77 \ \tilde C_{t\varphi}\tilde C_{\varphi G}\, .\label{eq:SMEFTLcorrectionLambda}
\end{align}
Note that this expression is not fully systematic. First, it retains only a part of the $\mathcal{O}(1/\Lambda^4)$-correction to the SM, since dimension-eight operators are not included, in contrast to the general form (\ref{eq:EWChLcorrectionLambda}). This can be improved by extending the relations of Table \ref{tab:SMEFTEWChL} to higher canonical dimensions. For example, dimension-eight contributions to (\ref{eq:SMEFTLcorrectionLambda}) can be found in \cite{Corbett:2021cil,Martin:2021cvs} (minor numerical differences arise due to a somewhat different treatment of higher order corrections). Second, it hides the possible implicit loop factor hidden in the coefficient $C_{\varphi G}$. A superficial inspection of (\ref{eq:SMEFTLcorrectionLambda}) would therefore lead to expect the highest deviations to be associated with this operator. As stated before, adding a consistent power-counting prescription for loops to the usual canonical counting in SMEFT can resolve this issue~\cite{Buchalla:2022vjp}. For instance, the coefficient in front of $\tilde C_{\varphi G}$ would change from $578.04$ to $3.6605$, which is a number
of order unity.

While in SMEFT the exact anomalous coupling between one Higgs boson and two gluons is given by an infinite tower of coefficients with increasing number of canonical dimension (and hence decreasing phenomenological importance), the formalism of the EWChL highlights the existence of a single anomalous coupling parameter $c_{ggh}$. It is therefore inconvenient to treat $C_{\varphi G}$ and its higher dimensional relatives $C_{\varphi G}^{(8)}$, $C_{\varphi G}^{(10)}$, etc. on unequal footing for single-Higgs processes. This also becomes clear in the context of QCD corrections. The latter can be summed up for all individual SMEFT contributions at once, which is equivalent to considering only one parameter from the start. Distinguishing contributions associated with different canonical dimensions is therefore not possible in the present case. For instance, $C_{\varphi G}$ and $C_{\varphi G}^{(8)}$ can not be extracted individually in single-Higgs processes, no matter how precise the experimental measurement is. However, processes involving two or more external Higgs states need additional coefficients, e.g. $c_{gghh}$.
In SMEFT at canonical dimension eight, $c_{ggh}$ and $c_{gghh}$ are represented by different linear combinations of $C_{\varphi G}$ and $ C_{\varphi G}^{(8)}$. Disentangling the latter coefficients thus requires the comparison of processes with varying number of Higgs particles~\cite{Buchalla:2018yce,Heinrich:2022idm}.


\section{Conclusions}
\label{sec:concl}

We have performed a detailed analysis of QCD corrections at NLO
for the Higgs-boson decays $h\to gg$ and $h\to\gamma\gamma$, allowing
for the presence of anomalous Higgs couplings from new-physics effects.
The natural framework for this task is provided by the EWChL, which
accounts for anomalous Higgs couplings at leading order in the EFT.
In addition, the EFT is governed by a power counting in loop orders,
which can be systematically combined with QCD perturbation theory.

For $h\to gg$ the relevant EFT coefficients are the local
Higgs-gluon coupling $c_{ggh}$, the Higgs-top coupling $c_t$ and,
to a lesser extent, the Higgs-bottom coupling $c_b$.
They are scale invariant under QCD.
No additional EFT parameters arise when the QCD calculation is
extended from LO to NLO.
For the $h\to gg$ rate the impact of QCD is known to be large,
with a K-factor of about $1.7$.
The uncertainties from scale dependence are reduced at NLO.
This also holds for the case of anomalous couplings,
in particular for the QCD coefficients of coupling factors,
such as $c^2_t$, $c_t c_{ggh}$ or $c^2_{ggh}$,
in the expression of the decay rate.
Those coefficients show NLO scale uncertainties at the $10\%$ level,
reduced by about a factor of 2 compared to LO.
QCD has less impact on $h\to\gamma\gamma$. In this case a NLO
treatment of QCD effects practically eliminates uncertainties
from perturbative QCD.

A new feature arising at NLO in QCD is that the analysis
becomes sensitive to ${\cal O}(\alpha_s)$ corrections
in the EFT coefficients $c_{ggh}$ and $c_{\gamma\gamma h}$.
Such terms are related to QCD corrections in the calculation of
these coefficients from matching to the underlying UV completion
of the EFT. In App.~\ref{app:toy} we have illustrated this with several toy models
for the UV physics.

We have also compared our treatment of the decays using the EWChL
with a description based on SMEFT.

The results presented here provide the basis for a consistent determination
of anomalous Higgs couplings from $h\to gg$ and $h\to\gamma\gamma$
at NLO in QCD.

\section*{Acknowledgements}
We thank Florian Pandler for useful discussions, and Michael Spira for valuable comments on the manuscript.
This work was supported by the Deutsche
Forschungsgemeinschaft (DFG, German Research Foundation) under grant
BU 1391/2-2 (project number 261324988) and by the DFG under
Germany’s Excellence Strategy – EXC-2094 – 390783311. 
Ch. M.-S. was supported in part by a Fellowship of the Studienstiftung
des deutschen Volkes (German Academic Scholarship Foundation).

\appendix


\section{IR subtraction}\label{app:IRsub}

Consider the NLO decay rate (\ref{eq:rateNLO_a}):
\begin{align}
    \Gamma_{h\to gg}^{NLO} &= \Gamma_{h\to gg}^{LO} + \Gamma_{h\to gg}^V + \Gamma_{h\to gg}^R\,.
\end{align}
The V and R corrections are separately IR divergent, with V containing explicit poles in $\epsilon$ coming from loop integrals carried out in dimensional regularisation, and R implicit phase space singularities related to soft or collinear final state particles. Summing both contributions, however, yields a finite result. In practice the phase space integrals cannot be evaluated analytically for most processes and numerical integration methods have to be applied. In this case, the pole cancellation cannot be checked directly. In order to still obtain sensible results, one has to systematically regulate the integrand in regions of the phase space where it diverges. As stated before, we adopt the antenna subtraction formalism~\cite{GehrmannDeRidder:2005cm,Daleo:2006xa,Currie:2013vh} for our setup, which proves to be particularly simple for the process at hand, i.e. $h\to gg$.

By construction the antenna subtraction term (S) reproduces the exact behaviour of the real correction matrix element in the IR singular limits for each colour level in the $1/N_c$ expansion individually. The difference of both can then be integrated numerically in a straightforward manner, as the integrand identically vanishes in all IR singular phase space regions. The subtraction term is constructed in such a way that, after appropriate factorization of the phase space, it can be integrated analytically over the phase space of the particle becoming unresolved. The integrated subtraction term (T) then exhibits explicit poles in $\epsilon$ which exactly cancel those of the virtual corrections. We have
\begin{align}
    \Gamma_{h\to gg}^V + \Gamma_{h\to gg}^R &= \underbrace{\int_2\mathrm{d}\Gamma_{h\to gg}^V-\int_2\mathrm{d}\Gamma_{h\to gg}^T}_{\mathrm{finite}} + \underbrace{\int_3\left(\mathrm{d}\Gamma_{h\to gg}^R-\mathrm{d}\Gamma_{h\to gg}^S\right)}_{\mathrm{finite}}\,,
\end{align}
where
\begin{align}
    \mathrm{d}\Gamma_{h\to gg}^T &= -\int_\mathrm{unres.}\mathrm{d}\Gamma_{h\to gg}^S\,.
\end{align}
The IR-subtracted expression for the $h\to ggg$  channel (\ref{eq:hggRggg}) is given by
\begin{align}
    \int_3\left(\mathrm{d}\Gamma_{h\to gg}^{R,ggg}\right.&\left. - \mathrm{d}\Gamma_{h\to gg}^{S,ggg}\right)\notag\\
        &= \frac{\alpha_s^3}{24\pi m_hv^2}N_c(N_c^2-1)\int\mathrm{d}\Phi_3\left(\sum_\lambda\mathcal{H}^\lambda\abs{A_{h\to ggg}^{(0)\lambda}}^2-2m_h^4F_3^0\abs{A_{h\to gg}^{(0)}}^2\right)\notag\\
        &= \frac{\alpha_s^3}{24\pi m_hv^2}N_c(N_c^2-1)\int\mathrm{d}\Phi_3\sum_\lambda\mathcal{H}^\lambda\left(\abs{A_{h\to ggg}^{(0)\lambda}}^2-\abs{A_{h\to gg}^{(0)}}^2\right)\,,
\end{align}
where our special kinematics ($1\to3$ decay) allows for the $F_3^0$ antenna function
(eq.~(7.8) in~\cite{GehrmannDeRidder:2005cm}) to be written as
\begin{align}
    m_h^4F_3^0 &= \frac{1}{2}\sum_\lambda\mathcal{H}^\lambda = \frac{m_h^8+s_{12}^4+s_{23}^4+s_{13}^4}{s_{12}s_{23}s_{13}}\,.
\end{align}
The IR-subtracted contribution of the $h\to gq\qb$ channel (\ref{eq:hggRgqqb})
to the rate evaluates to
\begin{align}
    \int_3\left(\mathrm{d}\Gamma_{h\to gg}^{R,gq\qb}\right.&\left. - \mathrm{d}\Gamma_{h\to gg}^{S,gq\qb}\right) \notag\\
        &= \frac{\alpha_s^3}{4\pi m_hv^2}(N_c^2-1)\int\mathrm{d}\Phi_3\left(\frac{s_{1q}^2+s_{1\qb}^2}{s_{q\qb}}\abs{A_{h\to gq\qb}^{(0)}}^2-m_h^4G_3^0\abs{A_{h\to gg}^{(0)}}^2\right)\notag\\
        &= \frac{\alpha_s^3}{4\pi m_hv^2}(N_c^2-1)\int\mathrm{d}\Phi_3\frac{s_{1q}^2+s_{1\qb}^2}{s_{q\qb}}\left(\abs{A_{h\to gq\qb}^{(0)}}^2-\abs{A_{h\to gg}^{(0)}}^2\right)\,,
\end{align}
where as before the $G_3^0$ antenna function
(eq.~(7.14) in~\cite{GehrmannDeRidder:2005cm}) can be simplified yielding
\begin{align}
    m_h^4G_3^0 &= \frac{s_{1q}^2+s_{1\qb}^2}{s_{q\qb}}\,.
\end{align}
Adding back the (now integrated) subtraction term to the virtual contribution
(\ref{eq:hggV}), we find
\begin{align}
    \Gamma_{h\to gg}^{V}-\Gamma_{h\to gg}^{T} &= \frac{\alpha_s^3}{512\pi^4}\frac{m_h^3}{v^2}(N_c^2-1)\left(\mathrm{Re}\left\{A_{h\to gg}^{(0)\dagger}A_{h\to gg}^{(1)}\right\}+2N_c\boldsymbol{J}_2^{(1)}\abs{A_{h\to gg}^{(0)}}^2\right)\,,
\end{align}
where the integrated antenna string is given by
\begin{align}
    \boldsymbol{J}_2^{(1)} &=  \boldsymbol{J}_2^{(1),ggg}+\frac{N_F}{N_c}\boldsymbol{J}_2^{(1),gq\qb} = \mu^{2\epsilon}\left(\frac{1}{3}\mathcal{F}_3^0(m_h^2)+\frac{N_F}{N_c}\mathcal{G}_3^0(m_h^2)\right)\,.
\end{align}
The integrated antenna functions $\mathcal{F}_3^0$ and $\mathcal{G}_3^0$ can be found in~\cite{GehrmannDeRidder:2005cm}. Explicitly,
\begin{align}
    \boldsymbol{J}_2^{(1)} &= -\frac{1}{2N_c}\mathrm{Re}\left\{N_cI(\epsilon)-\beta_0L+\frac{7}{3}N_F-\frac{73}{6}N_c\right\}+\order{\epsilon}\,,
\end{align}
where $L=\log\mu^2/m_h^2+i\pi$ and $I(\epsilon)$ has been defined
in~(\ref{eq:Iepsilon}). We thus find
\begin{multline}
    \Gamma_{h\to gg}^{V}-\Gamma_{h\to gg}^{T} = \frac{\alpha_s^3}{512\pi^4}\frac{m_h^3}{v^2}(N_c^2-1)\left(\left(\beta_0\log\frac{\mu^2}{m_h^2}-\frac{7}{3}N_F+\frac{73}{6}N_c\right)\abs{A_{h\to gg}^{(0)}}^2\right.\\  \left.+\frac{1}{2}\sum_qc_q\mathrm{Re}\left\{A_{h\to gg}^{(0)\dagger}A_{q,g,\mathrm{fin}}^{(1)}(\tau_q)\right\}\right)\,,
\end{multline}
which is finite as $\epsilon\to0$.


\section{Parametric suppression of the decay rate}\label{app:NNLOfix}
In a theory in which the coupling of the Higgs to gluons is given by an effective local interaction, such as the $c_{ggh}$ term in~(\ref{eq:ewchl}), the decay rate including $\order{\alpha_s}$ corrections can be given in exact form, as the integral over the phase space of the real radiation corrections can be calculated analytically. This is the case for the SM~\cite{Spira:1995rr}, considering only the top-quark and the limit $m_t\to\infty$, in which the top-quark loop is no longer resolved, generating an effective Higgs-gluon interaction of the form $hG_{\mu\nu}^aG^{a\mu\nu}$. The corresponding Wilson coefficient is known to N$^4$LO in QCD~\cite{Chetyrkin:2005ia,Schroder:2005hy,Spira:2016zna,Spira:2016ztx}. In this limit the relevant part of the effective Lagrangian~(\ref{eq:ewchl}) reduces to\footnote{See also Appendix~\ref{app:toy}. Note that here we explicitly set $N_c=3$.}
\begin{align}
  {\cal L}_{eff}^{\infty} &\supset \frac{\alpha_s}{8\pi}\left(c_{ggh}+
             \frac{2}{3}c_t\left(1+11\frac{\alpha_s}{4\pi}\right)\right)
             \frac{h}{v}G_{\mu\nu}^aG^{a\mu\nu}
           \label{eq:effL}\,,
\end{align}
where the superscript ``$\infty$'' indicates that we sent $m_t$ to infinity. Since the heavy top limit is actually a very good approximation already at LO~\cite{Spira:1995rr}, this form of the Lagrangian enables us to explore the interplay of the effective couplings $c_{ggh}$ and $c_t$ in a simple, yet not unrealistic scenario. The decay rate reads
\begin{align}
    \Gamma_{h\to gg}^\infty &= \Gamma_{h\to gg}^{LO,\infty}\left[1+\frac{\alpha_s}{\pi}\left(\mathcal{R}+\frac{11c_t}{3c_{ggh}+2c_t}\right)+\order{\alpha_s^2}\right]\,,
    \label{eq:NLO_ratemtinf}
\end{align}
with
\begin{align}
    \Gamma_{h\to gg}^{LO,\infty} &= \left(\frac{\alpha_s}{4\pi}\right)^2\frac{\left(3c_{ggh}+2c_t\right)^2}{18\pi}\frac{m_h^3}{v^2}\,,\label{eq:LO_ratemtinf1}
\end{align}
and
\begin{align}
    \mathcal{R} &= \frac{73}{4}-\frac{7N_F}{6}+\frac{33-2N_F}{6}\log\frac{\mu_R^2}{m_h^2}
\end{align}
is the finite contribution from the V and R corrections. The second term of the $\order{\alpha_s}$-correction is related to the $\order{\alpha_s}$-correction to the effective coupling, see~(\ref{eq:effL}). For later convenience, we define
\begin{align}
    \Delta &\coloneqq \frac{11c_t}{3c_{ggh}+2c_t}\,.
\end{align}
$c_{ggh}\approx-(2/3)c_t$ is obviously a critical region in the parameter space as the LO result (\ref{eq:LO_ratemtinf1}) becomes very small. While $\mathcal{R}>0$ for realistic $N_F$ and $\mu_R$ (i.e. $N_F=5$ and $\mu_R\approx m_h$), the term $\Delta$ can become large and negative, eventually rendering the whole NLO decay rate unphysical when $1+(\alpha_s/\pi)(\mathcal{R}+\Delta)<0$. This is an artefact of neglecting a certain part of the $\order{\alpha_s^2}$ corrections - (\ref{eq:NLO_ratemtinf}) is not a perfect square - as we will explain in the following.

For ${c_{ggh}=-(2/3)c_t}$, both the LO and NLO rate vanish identically. The Higgs gluon coupling becomes effectively $\order{\alpha_s^2}$,
\begin{align}
    \evalu{{\cal L}_{eff}^\infty}{c_{ggh}\to-\frac{2}{3}c_t} &\supset \frac{\alpha_s^2}{32\pi^2}\frac{22c_t}{3}\frac{h}{v}G_{\mu\nu}^aG^{a\mu\nu}\,,
\end{align}
and the rate starts at $\order{\alpha_s^4}$,
\begin{align}
    \evalu{\Gamma_{h\to gg}^\infty}{c_{ggh}=-\frac{2}{3}c_t} &= \left(\frac{\alpha_s}{4\pi}\right)^4\frac{242c_t^2}{9\pi}\frac{m_h^3}{v^2} + \order{\alpha_s^5}\,.\label{eq:app:NNLOfix}
\end{align}
This term is in fact a genuine part of the NNLO decay rate for arbitrary $c_{ggh}$ and $c_t$. While the LO and NLO parts as well as the other NNLO pieces of the rate are parametrically suppressed for $c_{ggh}\approx-(2/3)c_t$, this one is not. It will be the dominant contribution to the decay rate in this regime, and should therefore not be neglected, irrespective of being formally of higher order in perturbation theory. We will thus define the NLO rate to include~(\ref{eq:app:NNLOfix}),
\begin{align}
    \Gamma_{h\to gg}^{NLO,\infty} &= \Gamma_{h\to gg}^{LO,\infty}\left[1+\frac{\alpha_s}{\pi}\left(\mathcal{R}+\Delta\right)\right]+\left(\frac{\alpha_s}{4\pi}\right)^4\frac{242c_t^2}{9\pi}\frac{m_h^3}{v^2}\\
        &= \Gamma_{h\to gg}^{LO,\infty}\left[\left(1+\frac{\alpha_s}{\pi}\frac{\Delta}{2}\right)^2 + \frac{\alpha_s}{\pi}\mathcal{R}\right]\,.
\end{align}
$\Delta$ now only appears in a square and the rate will always be positive for positive $\mathcal{R}$. For $c_{ggh}\approx-(2/3)c_t$, we will thus obtain a more reliable prediction, while for all other cases, the contribution of the $\Delta^2$ term will be subdominant.

It turns out that the same parametric suppression of the LO and NLO rate occurs when we retain a finite top mass, namely in the region $c_{ggh} \approx -(c_t/2)A_q^{(0)}(\tau_t)$, see~(\ref{eq:Ahgg0}). This issue can again be resolved by adding~(\ref{eq:app:NNLOfix}), which in fact is nothing but the $m_t\to\infty$ limit of the suitably IR-regulated part of the virtual corrections squared contribution to the NNLO corrections\footnote{The double virtual corrections at the NNLO level consist of the interference of the Born-amplitude with the amplitude with two additional loops, but also of the square of the amplitude with one additional loop. The IR-divergences are compensated by double real and real virtual contributions.},
\begin{align}
     \evalu{\Gamma_{h\to gg}^{V\times V,\infty}}{\text{IR-regulated}} &= \left(\frac{\alpha_s}{4\pi}\right)^4\frac{c_t^2}{2\pi}\abs*{\lim_{\tau_t\to0}A_{q,g,\mathrm{fin}}^{(1)}(\tau_t)}^2\frac{m_h^3}{v^2}\\
        &= \left(\frac{\alpha_s}{4\pi}\right)^4\frac{242c_t^2}{9\pi}\frac{m_h^3}{v^2}\,,
\end{align}
with $A_{q,g,\mathrm{fin}}^{(1)}$ from equation~(\ref{eq:Aqgfin1}). We will settle for the heavy top limit here, since the mass effects in this perturbatively suppressed term will be negligible for most of the parameter space, but remark that if one wishes to explore the parametrically suppressed region, i.e. close to the vanishing LO rate, one should keep in mind that the calculation is effectively a LO calculation in this specific approximation.

In practice, this treatment amounts to a shift of the coefficient $A_{tt}^{NLO}$ in the rate as defined in~(\ref{eq:rateNLO}),
\begin{align}
    A_{tt}^{NLO} &\to A_{tt}^{NLO} + \delta A_{tt}^{NLO}\,,
\end{align}
with
\begin{align}
    \delta A_{tt}^{NLO} &= \left(\frac{\alpha_s}{4\pi}\right)^4\frac{242}{9\pi}\frac{m_h^3}{v^2}\,.
\end{align}
This shift is contained in all plots in Section~\ref{sec:pheno}, where also the bottom-quark contribution is included. Figure~\ref{fig:contour} serves as an empirical check that the latter does not introduce any new critical points; the global minimum of the rate is positive for arbitrary values of $c_{ggh}$, $c_t$ and $c_b$.

We remark that this is in general not true if we include further light quarks, such as the charm. Albeit we expect its contribution (for $c_c\sim\order{1}$) to be comparatively small due to its small mass, it will be possible to find particular combinations in the now four-dimensional parameter space spanned by $c_{ggh}$, $c_t$, $c_b$, $c_c$ for which the NLO rate will become negative, with very small magnitude. We checked for this particular scenario that it can only happen if $c_b$ and $c_c$ exceed $c_t$ and $c_{ggh}$ by at least one order of magnitude. Those configurations are well away from any region of phenomenological interest~\cite{deBlas:2018tjm}. In addition, the argument that the predictions in these cases are effectively of lower order in perturbation theory still holds, and thus they are subject to larger uncertainties.


\section{\boldmath Correlations of the parametric uncertainties of $\Gamma_{h\to gg}$ \unboldmath}
\label{app:correlations}
When calculating the parametric uncertainties of the decay rates eq.~(\ref{eq:rateLO}) or eq.~(\ref{eq:rateNLO}), one has to take into account the correlations between the respective uncertainties $\sigma_i$ of the coefficients $A_i$, as presented in Table~\ref{tab:coeffs}. Those correlations are described by the symmetric matrix $\rho$, so that
\begin{align}
    \sigma_\Gamma &= \sqrt{\sum_{i,k}\tilde{\sigma}_i\rho_{ik}\tilde{\sigma}_k} = \sqrt{\sum_i\tilde{\sigma}_i^2+\sum_{i\neq k}\tilde{\sigma}_i\rho_{ik}\tilde{\sigma}_k}\,,
\end{align}
where in case of the decay $h\to gg$ with anomalous couplings $c_{ggh}$, $c_t$ and $c_b$ the vector $\tilde{\sigma}$ is defined as
\begin{align}
    \tilde{\sigma} &= \left(c_{ggh}^2\sigma_{ggh},\, c_t^2\sigma_{tt},\, c_b^2\sigma_{bb},\, c_tc_{ggh}\sigma_{tg},\, c_bc_{ggh}\sigma_{bg},\, c_bc_t\sigma_{bt}\right)\,.
\end{align}
At LO and NLO in QCD the correlation matrices read
\begin{equation}
\rho^{LO} =
	\begin{pmatrix}
		1 & 1.000 & 0.351 & 1.000 & -0.735 & -0.736\\
		 1.000 & 1 & 0.350 & 1.000 & -0.734 & -0.735\\
		 0.351 & 0.350 & 1 & 0.350 & -0.893 & -0.892\\
		 1.000 & 1.000 & 0.350 & 1 & -0.735 & -0.735\\
		 -0.735 & -0.734 & -0.893 & -0.735 & 1 & 1.000\\
		 -0.736 & -0.735 & -0.892 & -0.735 & 1.000 & 1
	\end{pmatrix}\,,
\end{equation}
and
\begin{equation}
\rho^{NLO} =
	\begin{pmatrix}
         1 & 1.000 & 0.410 & 1.000 & -0.773 & -0.780\\
		 1.000 & 1 & 0.410 & 1.000 & -0.773 & -0.780\\
		 0.410 & 0.410 & 1 & 0.410 & -0.896 & -0.891\\
		 1.000 & 1.000 & 0.410 & 1 & -0.773 & -0.780\\
		 -0.773 & -0.773 & -0.896 & -0.773 & 1 & 1.000\\
		 -0.780 & -0.780 & -0.891 & -0.780 & 1.000 & 1
	\end{pmatrix}\,,
\end{equation}
respectively. At the accuracy we are working at, the impact of the shift~(\ref{eq:NNLOfix}) on $\rho^{NLO}$ is negligible.


\section{\boldmath Toy models for $c_{\gamma\gamma h}$ and
  $c_{ggh}$ \unboldmath}
\label{app:toy}

The Lagrangian~(\ref{eq:ewchl}) provides an effective description of the physics around the scale of electroweak symmetry breaking $v$. Short distance effects related to the scale $f\gg v$ are encoded into the Wilson coefficients such as $c_{ggh}$ and $c_{\gamma\gamma h}$. By experimentally constraining their values one can make statements about the characteristics of the unknown high scale physics, as different models of the UV theory lead to different predictions for the size of the effective couplings. In the following we will consider two toy models to sketch how the couplings $c_{ggh}$ and $c_{\gamma\gamma h}$ are related to the parameters of the full theory, including $\order{\alpha_s}$ corrections.

In both scenarios, we consider a new heavy particle, coloured and charged, mediating the coupling of the Higgs to photons and gluons through a loop, similar to the quark loops in the SM. Assuming that the interaction of the Higgs with the new particle $i$ can be implemented by the substitution
\begin{align}
    m_i &\to m_i\left(1+g_{i,h}\frac{h}{v}\right)\label{eq:gih}
\end{align}
in its mass term, with $g_{i,h}$ an arbitrary $\order{1}$ constant, the effective coupling can be derived by means of low-energy theorems (LETs)~\cite{Ellis:1975ap,Shifman:1979eb,Spira:1995rr,Kniehl:1995tn}, which have been used in the calculation of various Higgs production and decay modes~\cite{Kniehl:1994ju,Kniehl:1995gj,Spira:1995rr,Dawson:1996xz,Muhlleitner:2006wx}. Similar approaches~\cite{Kramer:1996iq,Chetyrkin:1997un,Chetyrkin:2005ia,Schroder:2005hy,Baikov:2006ch,Spira:2016zna} are related to decoupling relations~\cite{Bernreuther:1981sg,Larin:1994va}, which connect gauge couplings in the full theory to those in the effective theory with one or several degrees of freedom removed. The effective couplings can then be expressed~\cite{Chetyrkin:1997un} in terms of gauge independent objects like the \MSb $\beta$-functions for the strong and electromagnetic coupling, and the QCD anomalous mass dimension $\gamma_m$ of the heavy particle\footnote{In the original references, decoupling relations were investigated with a heavy quark in mind, but the results can be used for scalar-induced Higgs-gluon and Higgs-photon interactions, too, assuming (\ref{eq:gih}) holds.},
\begin{align}
     c_{ggh} &= -\frac{2\pi g_{i,h}}{\alpha_s^2\left(1-\gamma_m(\alpha_s')\right)}\left(\beta(\alpha_s)-\beta'(\alpha_s')\pdv[\alpha_s]{\alpha_s'}\right)\label{eq:C1decoupling}\,,\\
     c_{\gamma\gamma h} &= -\frac{2\pi g_{i,h}}{\alpha^2\left(1-\gamma_m(\alpha_s')\right)}\left(\beta_\gamma(\alpha,\alpha_s)-\beta'_\gamma(\alpha',\alpha_s')\pdv[\alpha]{\alpha'}-\beta'(\alpha_s')\pdv[\alpha]{\alpha_s'}\right)\label{eq:C1gamdecoupling}\,.
\end{align}
Quantities marked with a prime are to be evaluated in the full theory, i.e. including the heavy particle, while otherwise the effective theory without the heavy particle has to be employed. The pure QCD and mixed QED-QCD $\beta$-functions are defined through
\begin{align}
    \beta(\alpha_s) \equiv \mu\dv[\alpha_s]{\mu} &= -2\alpha_s\left[\epsilon+\left(\frac{\alpha_s}{4\pi}\right)\beta^0+\left(\frac{\alpha_s}{4\pi}\right)^2\beta^1+\order{\alpha_s^3}\right]
\end{align}
and
\begin{align}
    \beta_\gamma(\alpha,\alpha_s) \equiv \mu\dv[\alpha]{\mu} &= -2\alpha\left[\epsilon+\left(\frac{\alpha}{4\pi}\right)\beta_\gamma^0+\left(\frac{\alpha}{4\pi}\right)\left(\frac{\alpha_s}{4\pi}\right)\beta_\gamma^1+\order{\alpha^2,\alpha\alpha_s^2}\right]\,,
\end{align}
respectively. The anomalous mass dimension is given by
\begin{align}
    \gamma_m \equiv \frac{\mu}{m}\dv[m]{\mu} &= -\left(\frac{\alpha_s}{4\pi}\right)\gamma_m^0 + \order{\alpha_s^2}\,.
\end{align}
Expanding the expressions~(\ref{eq:C1decoupling}) and~(\ref{eq:C1gamdecoupling}) up to NLO in $\alpha_s$ we find
\begin{align}
    c_{ggh} &= -g_{i,h}\left[\Delta\beta^0 + \left(\frac{\alpha_s}{4\pi}\right)\left(\Delta\beta^1-\Delta\beta^0\gamma_m^0\right)\right] + \order{\alpha_s^2}\,,\label{eq:C1decoupling_exp}\\
    c_{\gamma\gamma h} &= -g_{i,h}\left[\Delta\beta_\gamma^0 + \left(\frac{\alpha_s}{4\pi}\right)\left(\Delta\beta_\gamma^1-\Delta\beta_\gamma^0\gamma_m^0\right)\right] + \order{\alpha_s^2}\label{eq:C1gamdecoupling_exp}\,,
\end{align}
where we defined
\begin{align}
    \Delta\beta^i &= \beta^{\prime i}-\beta^{i}\,, &   \Delta\beta_\gamma^i &= \beta_\gamma^{\prime i}-\beta_\gamma^{i}\,,
\end{align}
i.e. the difference of the $\beta$-function coefficients in the full and effective theory.

\subsection{Fermion}
As a first example, we can consider a model with an additional heavy fermion $F$ with electric charge $Q_F$ in an arbitrary representation $R$ of $SU(N_c)$:
\begin{align}
    \mathcal{L} &= \mathcal{L}_{\mathrm{SM}} + \bar{F}(i\slashed{D})F - m_F\left(1+g_{F,h}\frac{h}{v}\right)\bar{F}F\,.
\end{align}
We then find~\cite{Jones:1981we}
\begin{align}
    \Delta\beta^0 &= -\frac{4}{3}T_R\,,&
    \Delta\beta^1 &= -\left(\frac{20}{3}C_A+4C_2(R)\right)T_R\,,\\
    \Delta\beta_\gamma^0 &= -\frac{4}{3}d(R)Q_F^2\,,&
    \Delta\beta_\gamma^1 &= -4C_2(R)d(R)Q_F^2\,,\\
\end{align}
for the differences of $\beta$-function coefficients and
\begin{align}
    \gamma_m^0 &= 6C_2(R)\,,\label{eq:gammamF}
\end{align}
for the leading coefficient of the QCD anomalous mass dimension of the fermion, which is a known textbook result~\cite{Peskin:1995ev,Schwartz_2013}. We checked~(\ref{eq:gammamF}) by explicitly carrying out the mass renormalization of the fermion. Here $d(R)$, $C_2(R)$ and $T_R$ are the dimension, the quadratic Casimir and the Dynkin index of representation $R$, respectively. The latter defines the normalisation of the generators in the given group representation, $\Tr[T_R^aT_R^b]=T_R\delta^{ab}$. $C_A$ is the quadratic Casimir of the adjoint representation. Plugging these quantities into~(\ref{eq:C1decoupling_exp}) and~(\ref{eq:C1gamdecoupling_exp}) and omitting terms of $\order{\alpha_s^2}$, we obtain
\begin{align}
    c_{ggh} &= \frac{4}{3}g_{F,h}T_R\left[1+\left(\frac{\alpha_s}{4\pi}\right)\left(5C_A-3C_2(R)\right)\right]\,,\\
    c_{\gamma\gamma h} &= \frac{4}{3}g_{F,h}d(R)Q_F^2\left[1+\left(\frac{\alpha_s}{4\pi}\right)\left(-3C_2(R)\right)\right]\,.
\end{align}
For a fermion in the fundamental representation of $SU(N_c)$ with $N_c=3$ we have
\begin{align}
    c_{ggh} &= \frac{2}{3}g_{F,h}\left[1+\frac{11}{4}\frac{\alpha_s}{\pi}\right]\,,\\
    c_{\gamma\gamma h} &= 4g_{F,h}Q_F^2\left[1-\frac{\alpha_s}{\pi}\right]\,.
\end{align}
Of course, this result coincides with the heavy-top limit in the SM.

\subsection{Scalar}
We can also consider\footnote{Only after we finished the following calculation we became aware of~\cite{Gori:2013mia}, in which a very similar model has been considered, focusing on the Higgs-gluon coupling.} a scalar with electric charge $Q_S$ in representation $R$ of $SU(N_c)$,
\begin{align}
    \mathcal{L} &= \mathcal{L}_{\mathrm{SM}} + \abs*{D_\mu S}^2 -m_S^2S^* S\left[1+\sum_nf_n\left(\frac{h}{v}\right)^n\right] - g_s^2\lambda_S^{ijkl}S_i^* S_k^* S_jS_l\,,\label{eq:LagS}
\end{align}
where $i,j,k,l=1,\dots,d(R)$ are colour indices. The coupling constant for a single Higgs to the scalar according to~(\ref{eq:gih}) is then given by\footnote{Note that upon integrating out the scalar, the Lagrangian~(\ref{eq:LagS}) also generates effective couplings involving an arbitrary number of Higgs particles. The derivation of the corresponding coupling constants requires a generalisation of the expressions~(\ref{eq:C1decoupling}) and~(\ref{eq:C1gamdecoupling}), which is beyond the scope of this appendix. We will therefore restrict ourselves to the single Higgs case.}
\begin{align}
    g_{S,h} &= \frac{f_1}{2}\,.
\end{align}
In (\ref{eq:LagS}) we assumed that the quartic coupling is proportional to $g_s^2$, as is the case in some supersymmetric scenarios. The coupling $\lambda_S$ depends on the concrete model under consideration, but in any case obeys
\begin{align}
    \left(\lambda_S^{ijkl}\right)^* &= \lambda_S^{jilk}\,.
\end{align}
Furthermore it is symmetric wrt. $i\leftrightarrow k$ and $j\leftrightarrow l$. If we are interested in a model where the four-scalar interaction does not contribute to the QCD corrections, and consequently does not affect the Higgs-gluon and Higgs-photon effective couplings to $\order{\alpha_s}$, we can simply set $\lambda_S=0$ in the following results.

The $\beta$-function differences read~\cite{Jones:1981we}
\begin{align}
    \Delta\beta^0 &= -\frac{1}{3}T_R\,, &
    \Delta\beta^1 &= -\left(\frac{2}{3}C_A+4C_2(R)\right)T_R\,,\\
    \Delta\beta_\gamma^0 &= -\frac{1}{3}d(R)Q_S^2\,, &
    \Delta\beta_\gamma^1 &= -4C_2(R)d(R)Q_S^2\,.
\end{align}
The scalar self-interaction does not enter the $\beta$-functions before the three-loop level~\cite{Curtright:1979mg}, but appears in the LO QCD anomalous mass dimension of the scalar,
\begin{align}
    \gamma_m^0 &= 3C_2(R)-\frac{4}{d(R)}\sum_{i,k}\lambda_S^{iikk}\,.
\end{align}
We derived $\gamma_m^0$ by explicit calculation of the mass renormalisation of the scalar; our result agrees with~\cite{Gori:2013mia}. Plugging everything into~(\ref{eq:C1decoupling_exp}) and~(\ref{eq:C1gamdecoupling_exp}) we obtain (up to terms $\order{\alpha_s^2}$)
\begin{align}
    c_{ggh} &= \frac{g_{S,h}}{3}T_R\left[1+\left(\frac{\alpha_s}{4\pi}\right)\left(2C_A+9C_2(R)+\frac{4}{d(R)}\sum_{i,k}\lambda_S^{iikk}\right)\right]\,,\\
    c_{\gamma\gamma h} &= \frac{g_{S,h}}{3}d(R)Q_S^2\left[1+\left(\frac{\alpha_s}{4\pi}\right)\left(9C_2(R)+\frac{4}{d(R)}\sum_{i,k}\lambda_S^{iikk}\right)\right]\,.
\end{align}
For a scalar in the fundamental representation of $SU(N_c)$ with $N_c=3$ we have
\begin{align}
    c_{ggh} &= \frac{g_{S,h}}{6}\left[1+\frac{9}{2}\frac{\alpha_s}{\pi}\right]\,,\\
    c_{\gamma\gamma h} &= g_{S,h}Q_S^2\left[1+3\frac{\alpha_s}{\pi}\right]\,,
\end{align}
if the quartic interaction does not contribute. In case the scalar is a squark in the MSSM, again in the fundamental representation, the coupling $\lambda_S$ is given by~\cite{Bilal:2001nv,Shifman:2012zz}
\begin{align}
    \lambda_S^{ijkl} &= \frac{1}{4}\left(T^a_{ij}T^a_{kl}+T^a_{il}T^a_{kj}\right)   \qquad\Rightarrow\qquad \sum_{i,k}\lambda_S^{iikk} = \frac{1}{4}N_cC_F\,,
\end{align}
where $T^a$ are the $SU(N_c)$ generators in the fundamental representation. For the effective couplings we then find ($N_c=3$)
\begin{align}
    c_{ggh} &= \frac{g_{S,h}}{6}\left[1+\frac{29}{6}\frac{\alpha_s}{\pi}\right]\,,\\
    c_{\gamma\gamma h} &= g_{S,h}Q_S^2\left[1+\frac{10}{3}\frac{\alpha_s}{\pi}\right]\,.
\end{align}
Note that these results do not take any gluino exchange into account~\cite{Muhlleitner:2008yw}. Our expressions agree with those presented in~\cite{Muhlleitner:2006wx,Gori:2013mia,Spira:2016ztx}. We again find consistent results by matching the scalar-loop induced amplitudes for $h\to\gamma\gamma$~\cite{Aglietti:2006tp} and $h\to gg$~\cite{Anastasiou:2006hc} in the limit of infinite scalar mass to the amplitudes in the effective theory with local Higgs-photon and Higgs-gluon interactions\footnote{Reference~\cite{Aglietti:2006tp} does not include the quartic self-interactions into the calculation of $h\to\gamma\gamma$, while~\cite{Anastasiou:2006hc} presents the results for $h\to gg$ in the $m_S\to\infty$ limit both with and without it. For further details concerning the previous literature, see footnote~10 in~\cite{Gori:2013mia}.}.

\clearpage


\printbibliography

\end{document}